\documentclass[acmsmall]{acmart}
\AtBeginDocument{%
  }
    
\begin{document} 

%\title{Entangled via Mycorrhiza with “FungiSync”: Propagating Umwelt in Co-located Cyberdelic Mixed Reality Ritual}

\title{Entangling Like Mycorrhizae: Mixing Realities Through Touch in "FungiSync"}

% Mycorrhizal network propagating cross Plants
% propagating Umwelt through touch

\author{Botao Amber Hu}\authornote{Corresponding author; also serves as a visiting lecturer at the School of Design \& Innovation, China Academy of Art.}
\orcid{0000-0002-4504-0941}
\affiliation{%
  \institution{Reality Design Lab}
  \city{New York City}
  \country{USA}
  }
\affiliation{%
  \institution{University of Oxford}
  \city{Oxford}
  \country{UK}
  }
\email{botao@reality.design}

\author{Danlin Huang}
\orcid{0009-0009-7954-0803}
\affiliation{%
  \department{School of Design \& Innovation}
    \institution{China Academy of Art}
    \city{Hangzhou}
  \country{China}
  }
\email{danlinhuang0428@gmail.com}

\author{Yilan Elan Tao}
\orcid{0000-0003-1691-9727}
\affiliation{%
  \institution{Reality Design Lab}
  \city{New York City}
  \country{USA}
  }
\email{elan@reality.design}

\author{Xiaobo Aaron Hu}
\orcid{0009-0005-7124-2757}
\affiliation{%
  \institution{Independent}
  \city{Shanghai}
  \country{China}
  }
\email{agalloch21@gmail.com}

\author{Rem RunGu Lin}
\orcid{0000-0003-1931-7609}
\affiliation{%
  \institution{The Hong Kong University of Science and Technology (Guangzhou)}
  \city{Guangzhou}
  \country{China}
  }
\email{rlin408@connect.hkust-gz.edu.cn}

% Language Convey. 
% Embodied Sensoring. cannot convey. 大量的信息.
% 通过fungi的思考。
% 通过 non-verbal touch。
% human evolution 

\begin{abstract}
Mycorrhizal networks---often called nature's ``wood-wide web''---are vast underground mycelial systems that connect individual plants through countless hyphae of mycorrhizal fungi joining with plant roots. Through these hyphal webs, resources and signals---carbohydrates, minerals, and biochemical cues---are mutualistically exchanged and redistributed across plants, sustaining forests as relational symbiotic ecologies rather than isolated individuals. What is it like to be a plant within the wood-wide web? We present \emph{FungiSync}, a multi-person, co-located mixed reality (MR) experience that translates mycorrhizal interdependence into a felt, somaesthetic participatory ritual. Participants embody different forest plants by holding masquerade-style MR headset masks with wood-branch-like handles decorated with mushrooms. In MR, each participant perceives a distinct, audio-reactive psychedelic augmented reality overlay---composed of resource-representing visual elements---layered atop a shared physical terrain, symbolizing an individualized digital \emph{umwelt} (perceptual world). FungiSync reprograms human hand touch into a metaphorical mycorrhizal exchange. When participants touch hands, their digital \emph{umwelten} begin to entangle: visual elements leak, mix, and merge across perspectives, as if hyphae were forging new connections and carrying resources between hosts within a larger mycelial network. By making mycorrhizal interdependence perceptible through embodied contact, FungiSync invites participants to feel with \emph{fungal epistemics}---a more-than-human alternative way of knowing grounded in symbiotic relationality as both an aesthetic experience and an ethical orientation---offering a critique of the accelerated individualism characterizing our technology-mediated posthuman era.
\end{abstract}




\begin{teaserfigure}
    \centering
    \includegraphics[width=1\textwidth]{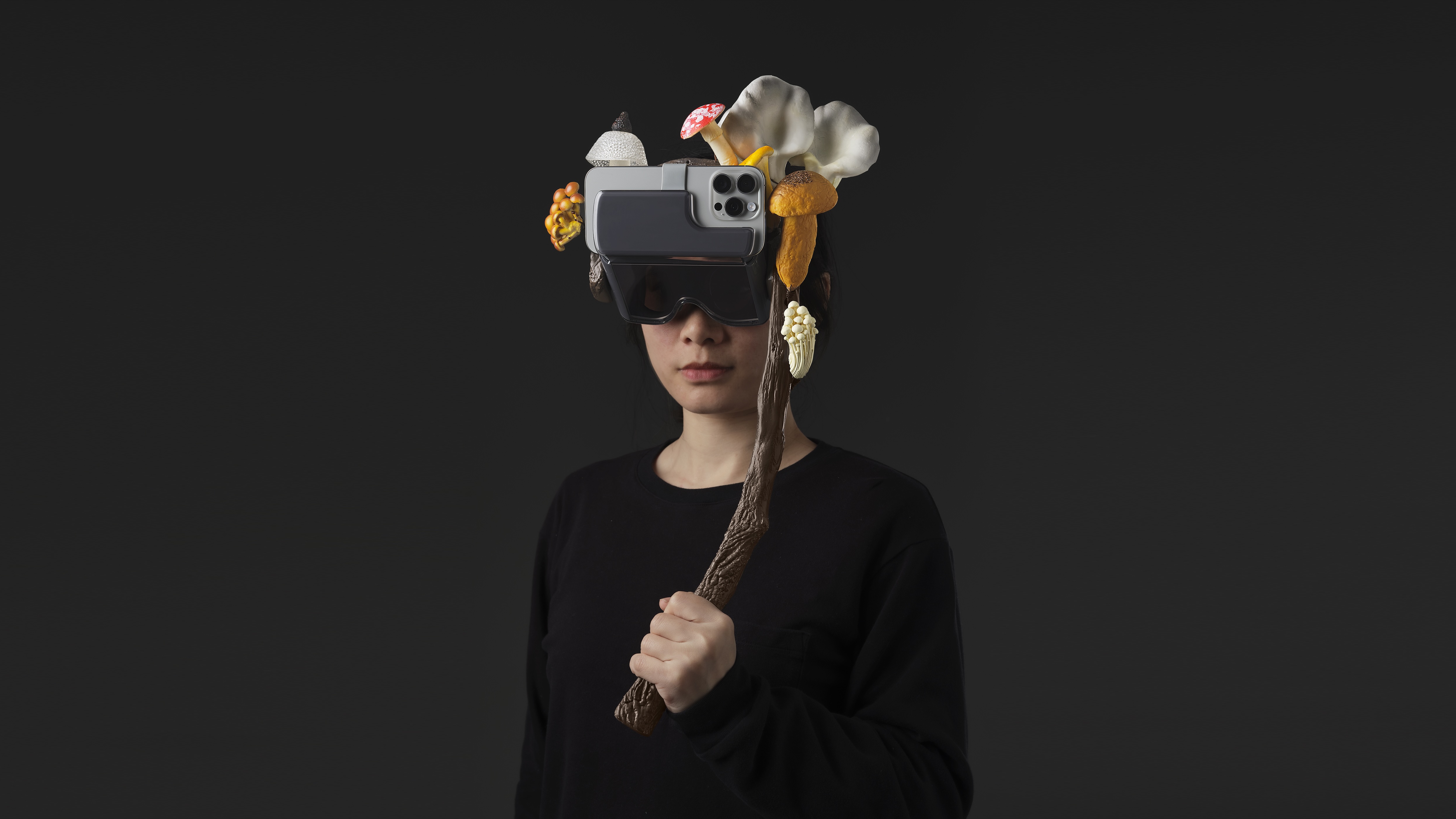}
    \caption{FungiSync is a mixed reality participatory ritual where co-located participants embody forest plants through handheld masquerade-style MR masks, each perceiving a unique psychedelic visual "umwelt" that merges and entangles with others through physical hand touch—embodying the mycorrhizal networks interconnecting plants as felt symbiosis.}
    \label{fig:teaser}
\end{teaserfigure}

\maketitle

% filter bubbles and ideological echo chambers
% algorithmic enclaves
% algorithmic politics
% https://www.caa-ins.org/archives/12472

% splinternet cyber-balkanization
% Virtual autism
% https://pmc.ncbi.nlm.nih.gov/articles/PMC5849631/
% “Early electronic screen exposure and autistic-like symptoms“
% Echo Chambers

% % solipsism
%% Cyberbalkanization
%% Mindblindness

\section{Introduction}

A mycorrhizal network is an underground system created by countless hyphae of mycorrhizal fungi interconnecting plant roots \cite{Simard2012Mycorrhizal}. These networks link individual plants in a shared web through which carbohydrates, minerals, water, and biochemical signals can be exchanged---a phenomenon popularly termed the ``wood-wide web'' \cite{sheldrake2021entangled}. Mycorrhizal relationships are typically mutualistic: shaded seedlings can receive carbohydrates from taller canopy trees capable of photosynthesis, while biochemical signals warning of herbivore attack can propagate through the network, enabling defensive responses across the forest \cite{Song2010Interplant}. This mycorrhizal wisdom emerges amid a broader ``fungal turn'' in contemporary humanities and culture, where fungi are increasingly mobilized as figures for distributed organization, multispecies relationality, and alternative political imaginaries \cite{Tsing2015Mushrooma, Bayley2025Mycelial}.

In parallel, contemporary digital life increasingly partitions perception into personalized feeds---worlds tuned by algorithmic selection and platform incentives. ``Filter bubble'' discourse has become shorthand for this condition: each person's informational environment becomes an individualized bubble, often opaque to others \cite{Pariser2011filter}. This resonates with \citet{Uexkull2010Forayb}'s concept of \emph{Umwelt}: each organism inhabits a species-specific perceptual world structured by what it can sense and act upon. If contemporary media systems produce individualized ``digital umwelten,'' then mycorrhizal networks offer a counter-image: not a single shared reality, but a relational ecology in which distinct worlds remain distinct while becoming mutually consequential through contact and exchange.

What is it like to be a tree in such a wood-wide web? In response, we present \textbf{FungiSync}, a multi-person, co-located mixed reality ritual that translates mycorrhizal interdependence into a \emph{felt} experience of entanglement. It is an open-ended performance where any participant can join by picking up and holding a masquerade-style MR headset mask with a wood-branch-like handle decorated with mushrooms. Each participant is randomly assigned an individualized, audio-reactive ``cyberdelic'' augmented reality overlay representing a plant's Umwelt, with unique visual elements attached to environmental surfaces representing resources the plant can sense and possess: water, sunlight, insects, minerals, sugar. We then \emph{reprogram} the human social protocol of hand touch into a metaphorical mycelial exchange: when two participants' hands touch, their respective MR overlays leak, merge, and swap visual elements across perspectives. Participants can walk and dance with different people to glimpse other umwelten. Their umwelten become abundant after exchanges but decay without sufficient connection.

The ritual is grounded in \citet{Shusterman2008Body}'s somaesthetics, treating bodily attunement and sensation as primary sites of meaning-making and ethical cultivation, and builds on \citet{Hook2018Designingf}'s soma design work translating somaesthetics into design practices. Through this entangling ritual, FungiSync proposes mycorrhiza not only as an ecological phenomenon but as an aesthetic-ethical orientation. We explore how participants can gain \emph{fungal epistemics} \cite{Bayley2025Mycelial} awareness through embodied ritual, and how such experience might critique the accelerated individualism of posthuman digital life by making interdependence tangible.

Our contributions are: (1) an MR art performance system that embodies mycorrhizal-relationality via multi-user somatic touch ritual to cultivate fungal epistemics as both aesthetic experience and ethical orientation; and (2) a ritual performance investigating how distinct ``digital umwelten'' can become entangled yet not homogenized, offering a more-than-human alternative grounded in symbiotic relationality that critiques the accelerated individualism of our technology-mediated era.

\section{Background and Related Work}
\subsection{Fungal Epistemics and Mycoaesthetics}

The scientific study of mycorrhizal networks has revealed that forests function as interconnected superorganisms rather than collections of competing individuals \cite{Simard2012Mycorrhizal}. Beyond biology, fungi have become vibrant symbols in humanities and arts discourse. Scholars speak of a ``fungal turn'' in contemporary thought, where fungal metaphors are invoked to rethink networks, community, and resilience in crisis. \citet{Tsing2015Mushrooma}'s ethnographic work on matsutake mushrooms articulates how fungi exemplify survival in capitalist ruins through collaborative, multispecies entanglements rather than individual optimization. \citet{sheldrake2021entangled}'s \emph{Entangled Life} has brought fungal thinking to popular audiences, emphasizing how mycelia challenge conventional notions of individuality, intelligence, and agency.

In aesthetic domains, \citet{cecire2024mycoaesthetics} define \emph{mycoaesthetics} as the array of cultural expressions that use fungi to sense and articulate ecological crisis. They observe that fungal forms have become ``culturally persuasive'' figures across design, fiction, and popular media, often standing for distributed connectivity or decomposition as creative forces. \citet{Bayley2025Mycelial} extends this discussion to \emph{fungal epistemics}, examining how ``mycelial thinking'' infiltrates theory and art to challenge anthropocentrism and emphasize relational ontologies. This ``fungal turn'' intersects with broader movements in feminist science studies and new materialism. \citet{Haraway2016Stayinga}'s concept of ``sympoiesis''---making-with rather than self-making---draws explicitly on symbiotic relationships including mycorrhizae to argue for a relational ethics of becoming-with other species.

FungiSync takes inspiration from these humanities perspectives, using art to explore what multispecies symbiosis might feel like and what it might teach us about being less individually bounded. Prior art projects have explored similar fungal themes in interactive contexts. \citet{Ji2025We}'s "We Are Entanglement" created a multi-user projection mapping experience visualizing participants as nodes in a mycelium network.  \citet{spacal2015connections}'s bioart installations integrate living mycelium with electronic sensors, creating hybrid organisms that respond to human presence.

\subsection{Ritual Design and Somaesthetics}

Our design process is deeply informed by somaesthetics, the philosophy and practice of cultivating embodied experience, as introduced by \citet{Shusterman2008Body}. Somaesthetics positions bodily awareness as a site of knowledge, agency, and ethical cultivation. In HCI, this has translated into \emph{soma design} methods that foreground first-person felt experience, movement, and sensory refinement as central to design \cite{Hook2018Designingf}.

Ritual theory further informs our approach. \citet{Meltzer1968INTERACTION}'s foundational work on interaction rituals examined how face-to-face encounters are structured by shared attention and emotional entrainment, producing social bonds and moral order through embodied performance. \citet{Collins2014Interaction} in \emph{"Interaction Ritual Chains"} proposed that successful rituals create symbols of group membership and energize individuals with emotional energy, while failed rituals deplete it.  
\citet{Yue2025Ritual} have analyzed ritual design in the digital age, identifying how contemporary artists and designers adapt ritual structures for technology-mediated contexts. \citet{Shildrick2002Embodyingc}'s work on embodiment and vulnerability further informs our approach, examining how encounters with other bodies can transform self-understanding.

Research on mediated touch and social touch technologies also informs our design. \citet{Haans2006Mediated} demonstrated that mediated touch can convey emotional content and enhance social presence even without visual or auditory cues. \citet{Price2022} found that novel touch interfaces can engender feelings of closeness and social presence. These precedents support FungiSync's central design move: using physical touch as the trigger for perceptual entanglement, thereby grounding an abstract ecological concept in the immediacy of bodily contact.

\subsection{Umwelt Theory and Digital Individualism}

The concept of \emph{Umwelt}, introduced by \citet{Uexkull2010Forayb} in his biosemiotic studies, provides theoretical foundation for FungiSync's approach to individualized perceptual worlds. Uexküll argued that each organism inhabits a species-specific perceptual world structured by its particular sensory capacities and action possibilities. There is no neutral, objective world that all organisms share; each species constructs its reality through the filter of its perceptual apparatus. This concept has been taken up in philosophy of mind, particularly in discussions of animal consciousness. \citet{Nagel1974Whata}'s famous question ``What is it like to be a bat?'' implicitly engages with umwelt thinking, acknowledging that other organisms may have subjective experiences radically unlike our own.

In digital media contexts, the umwelt concept gains new relevance. \citet{Pariser2011filter}'s analysis of ``filter bubbles'' describes how algorithmic personalization creates individualized information environments, effectively producing distinct digital umwelten for each user. \citet{Turkle2011Alone}'s \textit{Alone Together} extends this critique, examining how digital technologies paradoxically isolate individuals even while connecting them—we inhabit shared platforms yet experience increasingly privatized perceptual worlds, tethered to screens rather than to one another.

\subsection{Intercorporeal Design in Mixed Reality Performance}

Our work is inspired by emerging research on intercorporeal design in Extended Reality (XR). Intercorporeality refers to the interconnectedness of bodies---the idea that bodily experience is always shaped by the presence and actions of other bodies, a concept stemming from \citet{merleau2013phenomenology}'s phenomenology of perception. \citet{stepanova2024intercorporeal} have argued that XR systems can dissolve the self-other boundary, enabling what they call intercorporeal experiences where participants feel a sense of shared or merged embodiment. Several precedent artworks have explored intercorporeality in XR contexts. \citet{desnoyers2020body}'s \emph{Body RemiXer} connected an immersant wearing a VR headset with co-located participants interacting via projected visuals, creating asymmetric but complementary perceptions of a shared dance. \citet{lin2024cell}'s \emph{Cell Space} integrated MR headsets, neurofeedback, and contact improvisation dance, creating a system where participants' brain states influenced shared visual environments \cite{lin2024cell}. \citet{Oliveira2016}'s \emph{Machine to Be Another} enables embodied perspective-taking: two participants swap first-person views via VR cameras and synchronize movements to induce the illusion of inhabiting each other's body. FungiSync extends these explorations by grounding intercorporeal exchange in a specific ecological metaphor---mycorrhizal connection.

 \begin{figure}[ht!]
     \centering
     \includegraphics[width=0.95\linewidth]{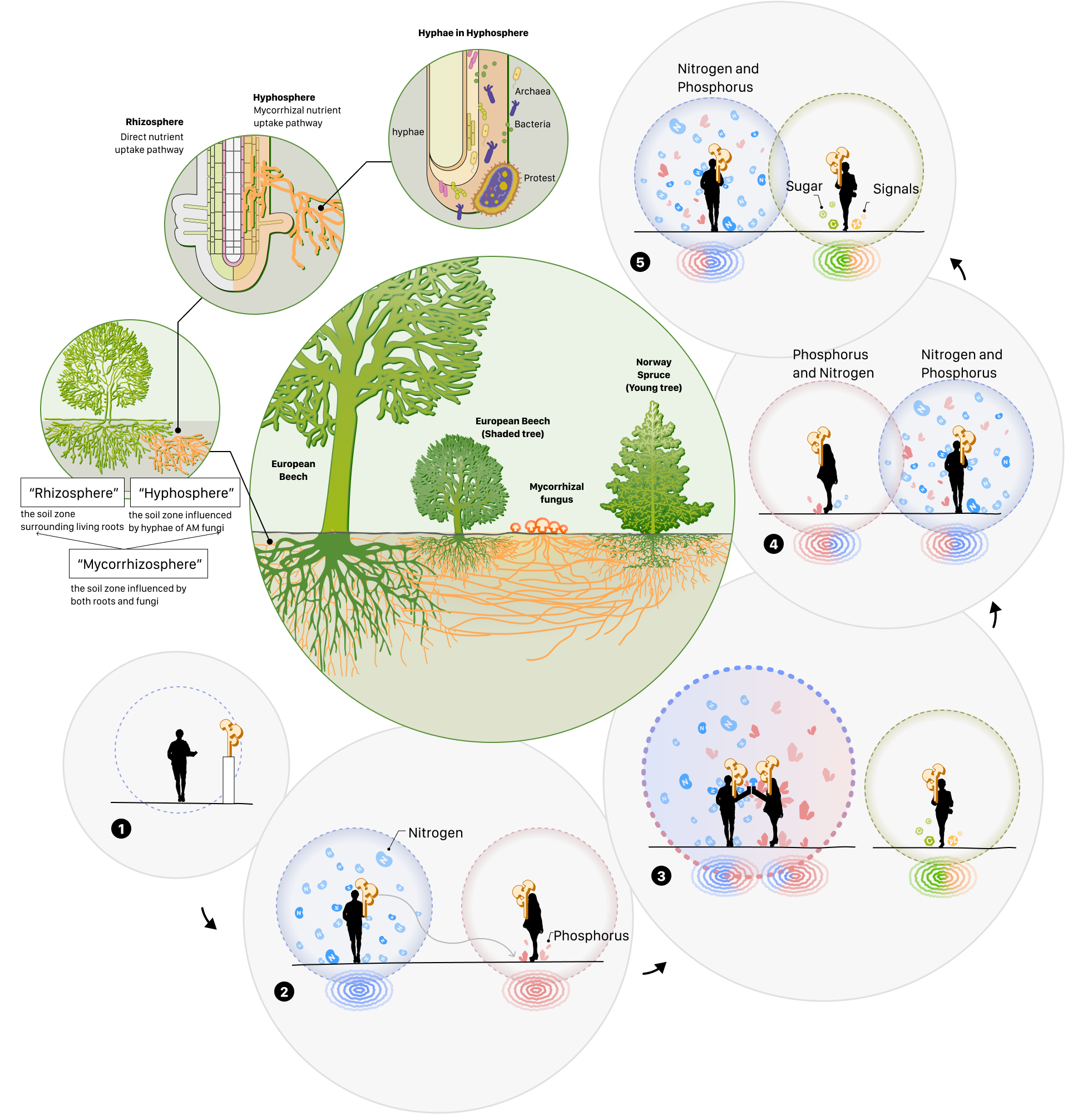}
     \caption{Conceptual framework mapping mycorrhizal network dynamics onto FungiSync's mixed reality interaction design. Center: The mycorrhizal network connecting forest plants (European Beech, Norway Spruce) through underground fungal hyphae, illustrating the rhizosphere, hyphosphere, and mycorrhizosphere zones. Sequence (1-5): The corresponding participant experience—(1) individual entering with isolated umwelt; (2) two participants with distinct perceptual worlds (nitrogen/phosphorus as visual elements); (3) physical hand touch mixing overlapping umwelten, triggering resource/signal exchange; (4) post-contact state showing transferred visual elements between participants; (5) enriched individual umwelten can continue to share elements with others, though elements gradually decay over time. The framework translates mycorrhizal mutualism—where plants exchange nutrients through fungal intermediaries—into a somaesthetic ritual where touch catalyzes perceptual entanglement, making invisible ecological interdependence viscerally felt.}
     \label{fig:concept}
 \end{figure}
 
\section{Conceptual Framework}

The core design concept of FungiSync is to let people embody entanglement in analogy to how plants experience mycorrhizal networks. Rather than explaining this phenomenon didactically, we focus on first-person enactment: what it might feel like, as an individual tree, to become interconnected via an unseen fungal medium. Drawing on somaesthetics, we treat the body as the primary site of knowledge production—understanding entanglement requires feeling it \cite{Shusterman2008Body}.

As Figure \ref{fig:concept} illustrates, just as European Beech and Norway Spruce connect through underground fungal hyphae—forming distinct rhizosphere, hyphosphere, and mycorrhizosphere zones—participants receive different visual "umwelten" based on their state: those without a mask exist within a limited rhizosphere, while those wearing the mask enter the mycorrhizosphere, their perception entangled with the fungal network like roots fused with mycorrhizal fungi. Each participant maintains a unique perspective, perceiving visual elements distinctive to their individual umwelt (nitrogen, phosphorus, sugar, signals, or water).

These distinct umwelten can mix through physical hand contact, functioning as a hyphal graft. The handshake—a human greeting ritual symbolizing trust and mutual acknowledgment \cite{Meltzer1968INTERACTION}—is redesigned to align with anastomosis, the biological process where compatible fungal hyphae fuse, creating continuous cytoplasmic connections through which resources flow \cite{Simard2012Mycorrhizal}. Hand touch temporarily entangles participants' perspectives, and this entanglement is bodily negotiated—participants decide when to connect and for how long, with visuals gradually reverting after disconnecting. Upon release, mixed elements slowly fade, returning each participant toward their original umwelt—but with traces lingering. Just as mycorrhizal organisms continuously adjust resource flows and feedback, participants must actively manage the rhythm of their connections to shape collective experience.

The ritual interaction sequence (Steps 1-5) demonstrates this process: (1) an individual enters with an isolated umwelt; (2) two participants perceive distinct visual worlds containing different elements; (3) hand touch triggers mixing, as AR overlays "leak" into each other—visual elements appear in the other's view, initially at the contact point and gradually spreading, with longer touch producing more complete mixing; (4) post-contact, transferred elements persist in both views; (5) enriched umwelten can share with new participants, though elements gradually decay. This models mycorrhizal networks, where connection is active and metabolically sustained rather than static. 

\section{System Design}

\subsection{Designing The Mask: Embodying Plants-with-Fungi}

FungiSync is built on accessible MR hardware to encourage co-located participation. We utilize smartphone-based optical see-through headsets called HoloKit X \cite{hu2024holokit}, an open-source toolkit that transforms an iPhone into an affordable AR headset by means of a stereoscopic visor and lens system, democratizing MR access through widely available components.

\begin{figure}[ht]
    \centering
        \includegraphics[width=0.3\linewidth]{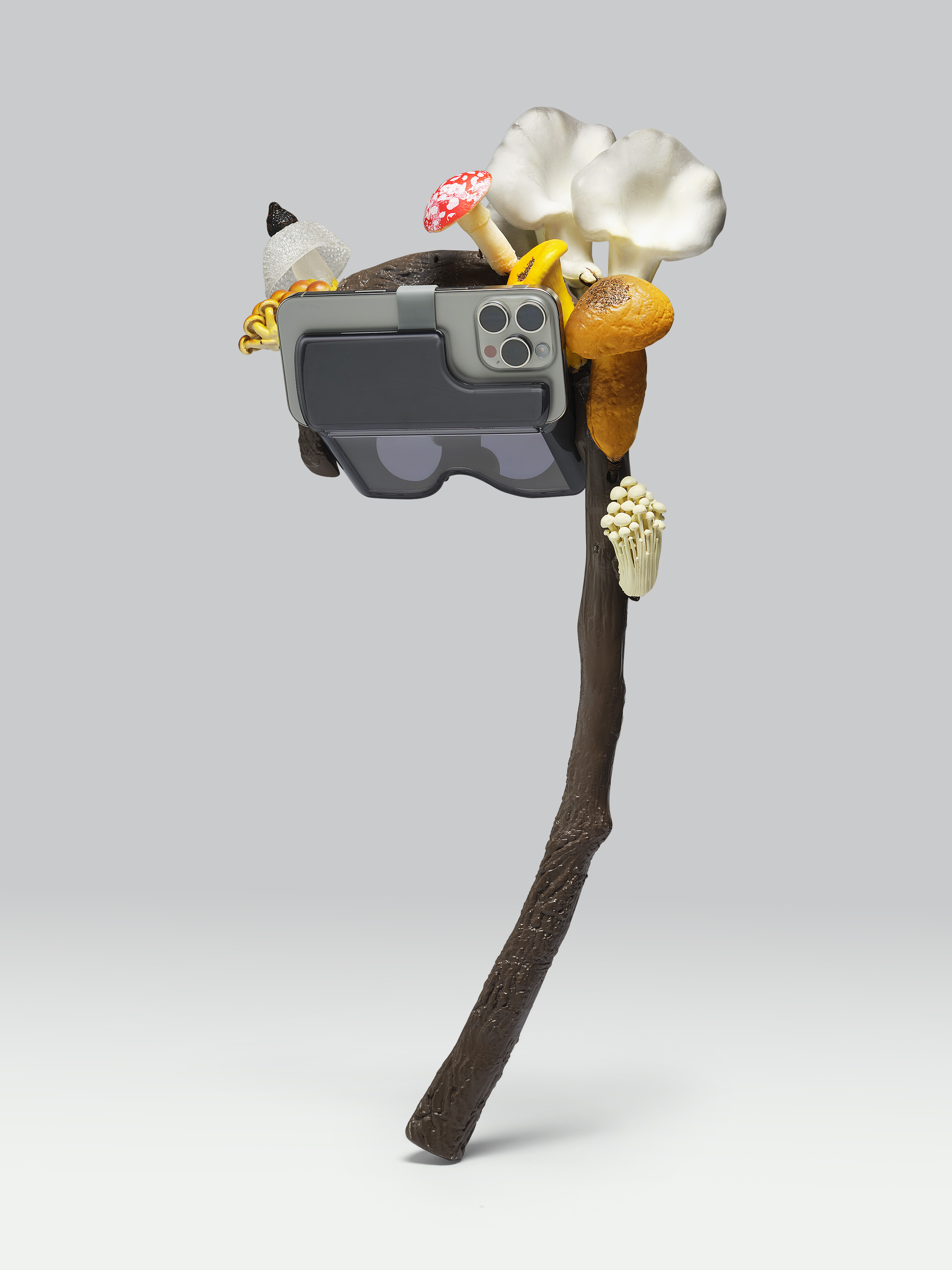}
        \includegraphics[width=0.63\linewidth]{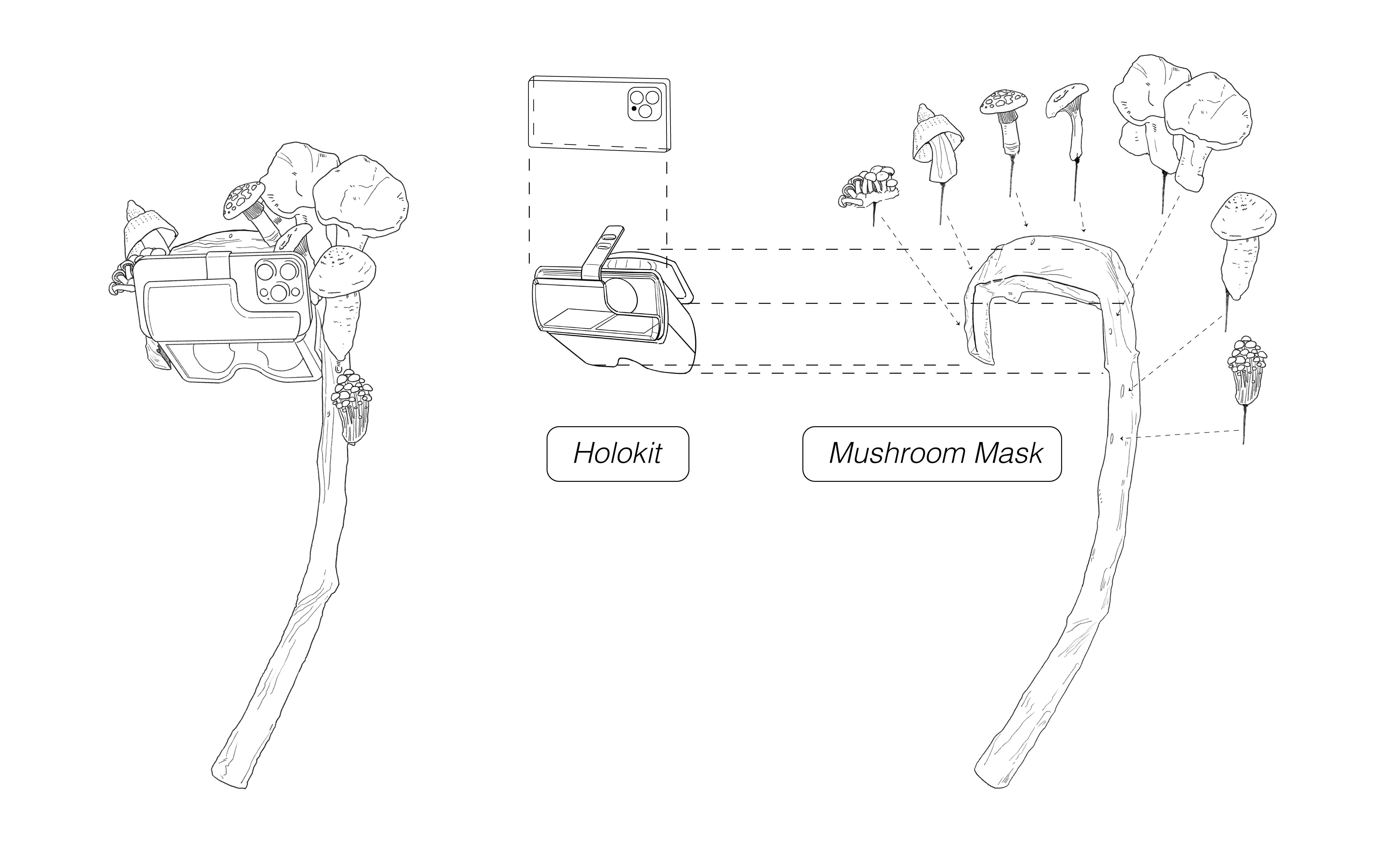}
    \caption{FungiSync mask design: a custom masquerade-style enclosure for the HoloKit X headset, featuring bark texture and mushroom caps to evoke a forest spirit colonized by fungi.}
    \label{fig:design-of-mask}
\end{figure}

We encase each HoloKit device within a custom-crafted masquerade-style mask, drawing inspiration from Venetian masquerade traditions that use masks to enable temporary identity transformation and social permission for unusual behavior. FungiSync's masks serve a similar function: they signal that participants have entered a special frame where normal social rules are suspended and new forms of ritual are invited. The mask provides a one-handed grip—a branch-like handle—so participants can easily hold the device to their face and lower it as needed, resolving ergonomic challenges while leaving one hand free for touching others. Figure \ref{fig:design-of-mask} illustrates the design components of the mask.

\begin{figure}[ht]
    \centering
    \includegraphics[width=0.9\linewidth]{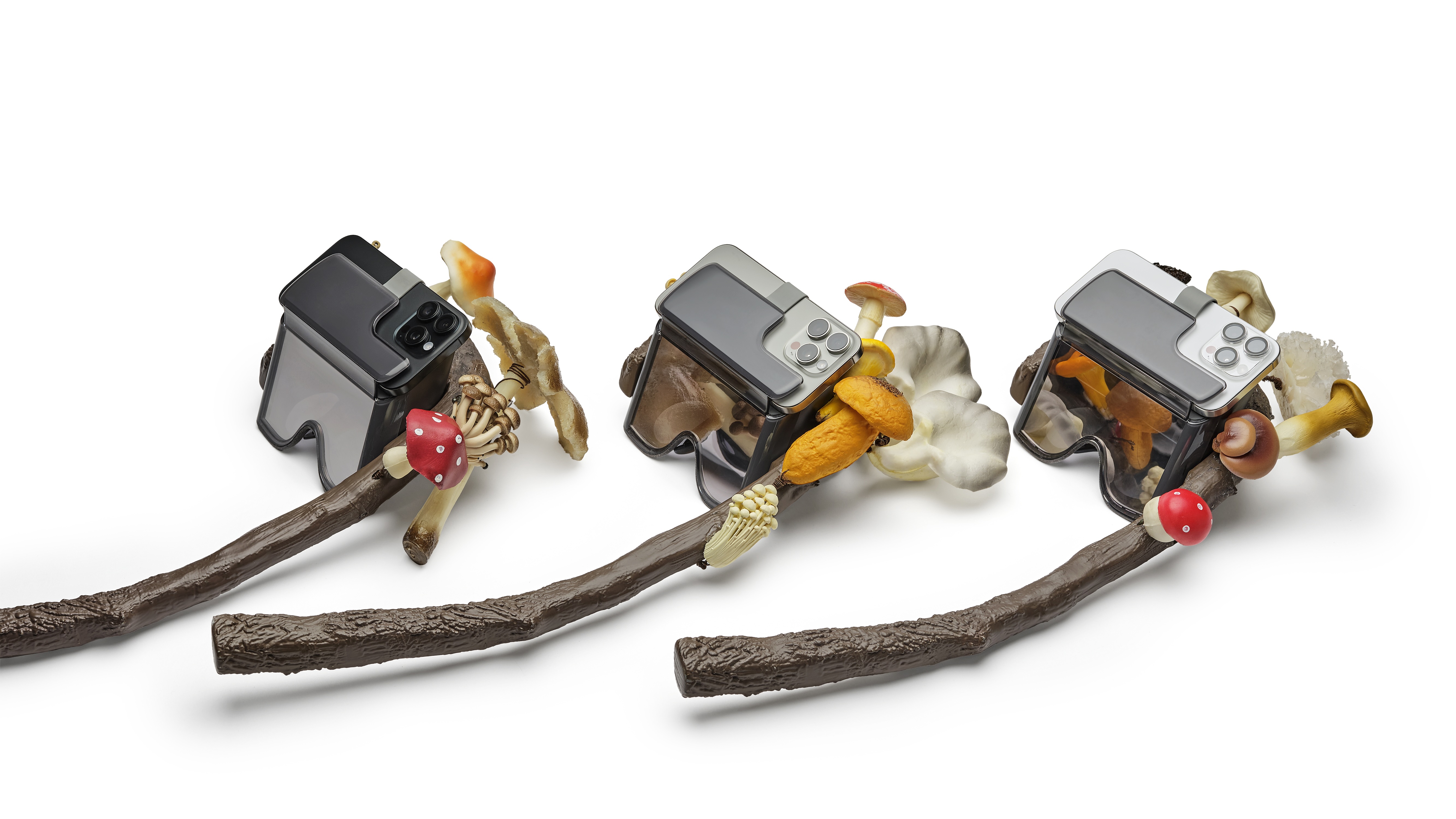}
        \includegraphics[width=0.45\linewidth]{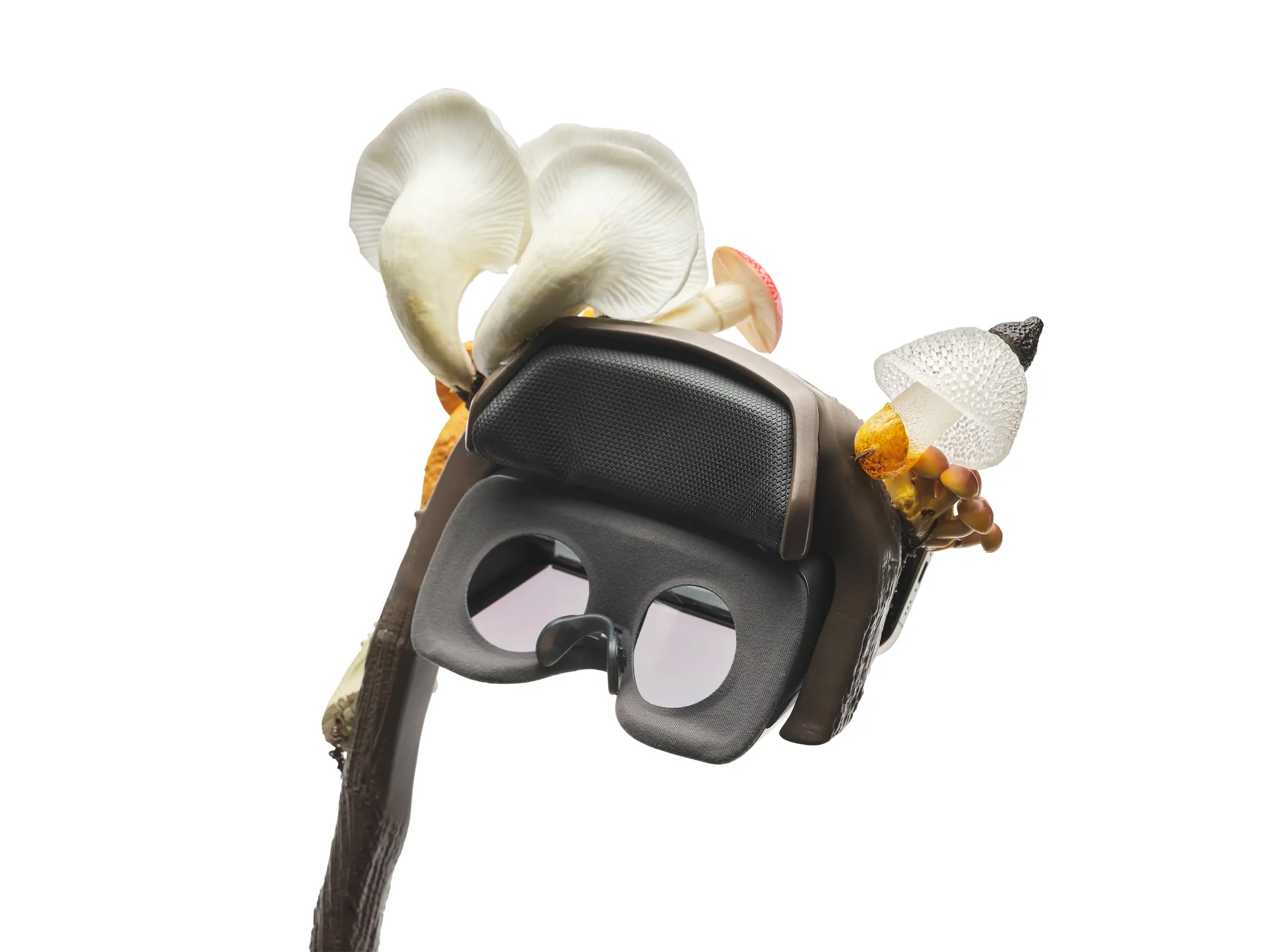}
        \includegraphics[width=0.45\linewidth]{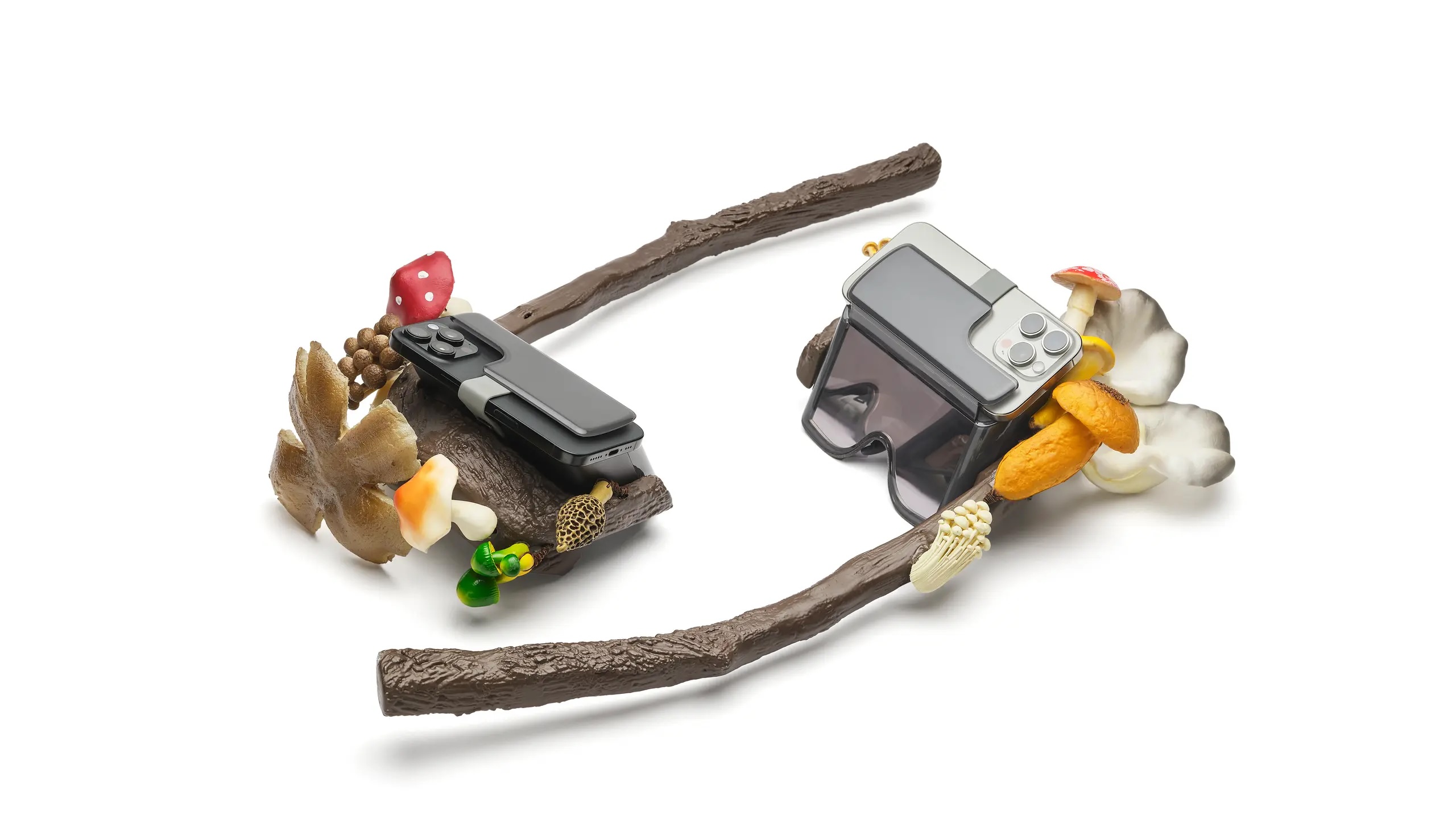}
    \caption{The Design of FungiSync Mask}
    \label{fig:masks_design}
\end{figure}

As Figure \ref{fig:masks_design} shows, the mask's design evokes a forest aesthetic: the shell is shaped and painted like a section of wood with bark texture, adorned with ready-made mushroom caps sprouting from the top and side, creating the impression that the mask itself has been colonized by fungi. The handle was modeled in CAD and 3D-printed in resin, then hand-painted to resemble a gnarled twig. This decorative motif externalizes the otherwise invisible fungal partner—holding the mask symbolizes that participants have become plants colonized by mycorrhizal fungi, ready to connect through the network.

\subsection{Designing the Umwelten: Cyberdelic Perceptual Worlds}

Each participant's default experience in FungiSync is to inhabit a distinct perceptual world represented by a custom visual augmented overlay. We developed MR overlays featuring abstract, dynamic, and audio-reactive visual elements—nitrogen, phosphorus, sugar, signals, or water—as poetic proxies for the kinds of resources and energies that flow through a forest. Each participant perceives only their assigned element, evoking a plant's umwelt: a limited sensory world defined by what that organism can receive and process.

The aesthetic is deliberately \textit{cyberdelic}, a term merging cybernetics with the psychedelic \cite{smith2022cyberdelics}. We adopted this style to suggest an altered state of consciousness, appropriate for an experience that is part science metaphor and part ritual. The visual language draws on the phenomenology of psychedelic experience—particularly the distortion of ordinary reality and heightened awareness of hidden patterns. This approach aligns with our goal of embodying plants with fungi: participants perceive their environment through unique psychedelic illusions, each representing a distinct umwelt—a perceptual world as alien and partial as, for example, that of a tree sensing chemical flows beneath the soil.

\begin{figure}
    \centering
    \includegraphics[width=0.9\linewidth]{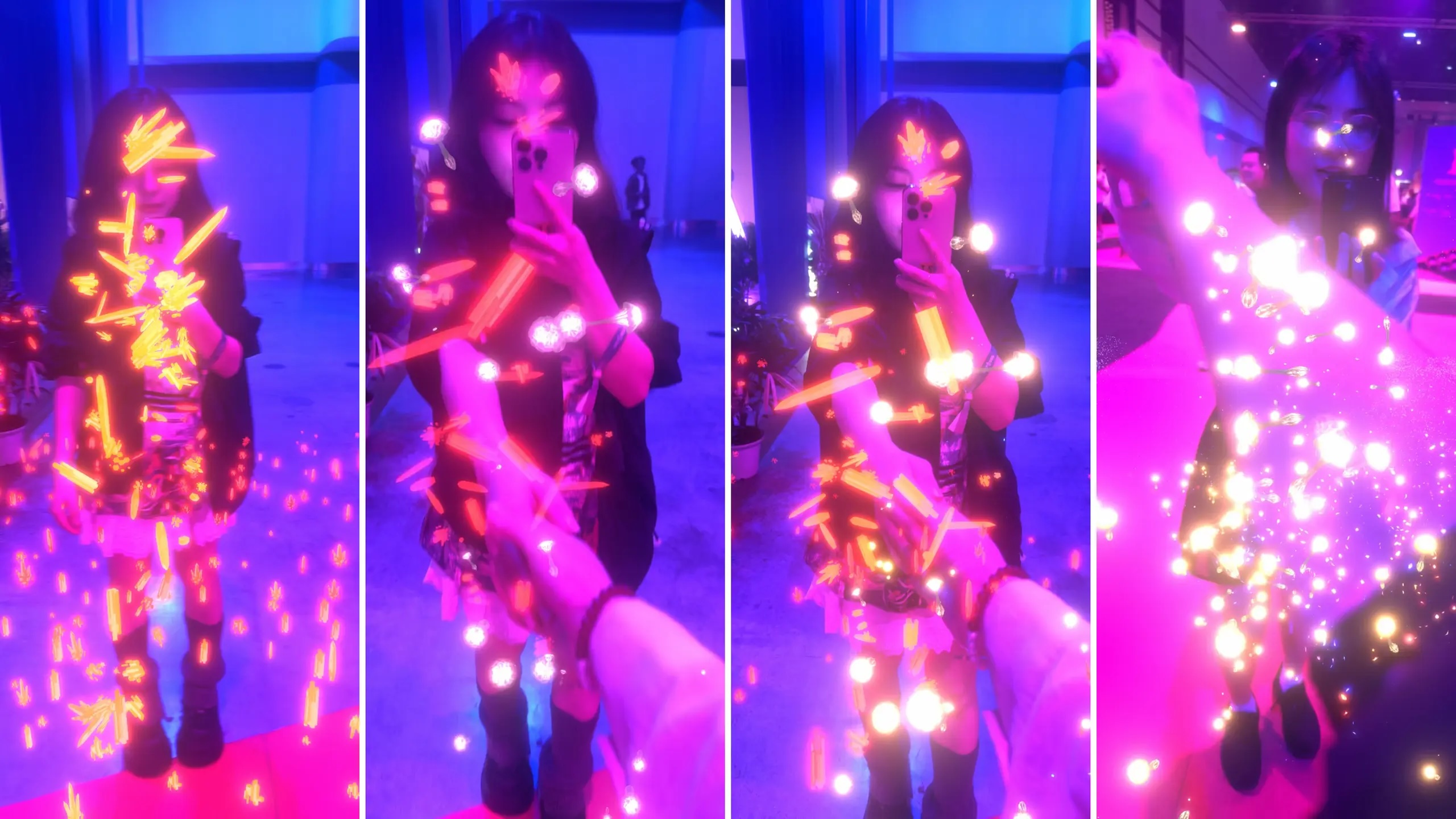}
    \caption{The cyberdelic MR visuals mix between participants when hand touch triggers connection, simulating resource exchange through mycorrhizal networks.}
    \label{fig:bodycontact2}
\end{figure}

As Figure \ref{fig:bodycontact2} shows, the visual layers are implemented using Unity's Visual Effect Graph and custom shaders. The iPhone's LiDAR sensor provides a real-time 3D mesh of the environment for occlusion and surface attachment, allowing virtual elements to appear as if growing on physical surfaces and enhancing the sense that MR content is embedded in reality. Each elements is also audio-reactive: microphone input is fed into a spectrum analyzer, and the resulting frequency data drives visual parameters. If one participant claps or speaks, the MR overlays of all participants ripple in their own characteristic ways, creating subtle background interdependence even before any physical touch occurs.

\begin{figure}
    \centering
    \includegraphics[width=0.9\linewidth]{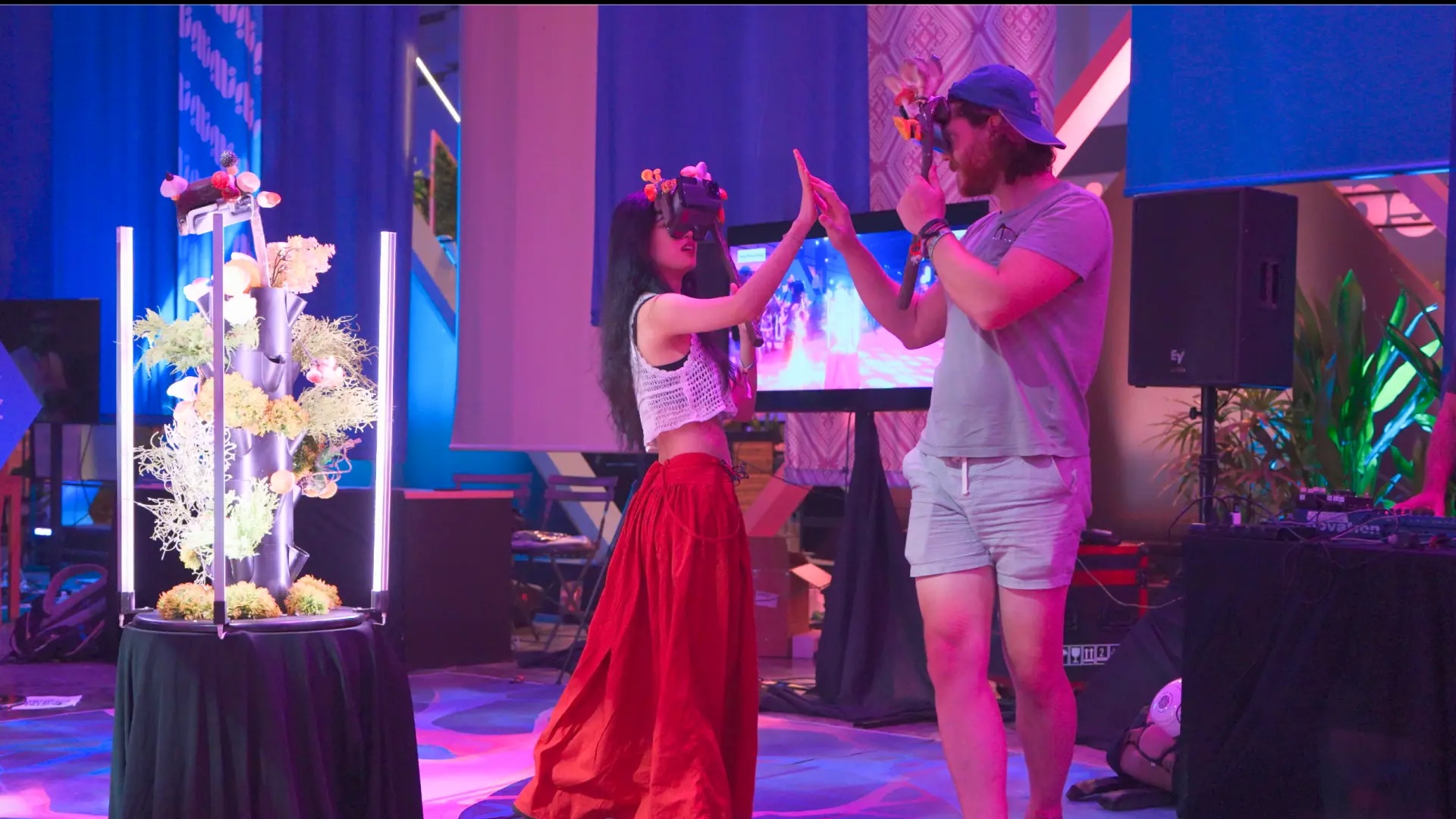}
    
    \caption{Hand touch between participants triggers the mixing of cyberdelic MR overlays, simulating mycorrhizal resource exchange through fungal hyphae.}
    \label{fig:bodycontact}
\end{figure}

\begin{figure}
    \centering
    \includegraphics[width=0.9\linewidth]{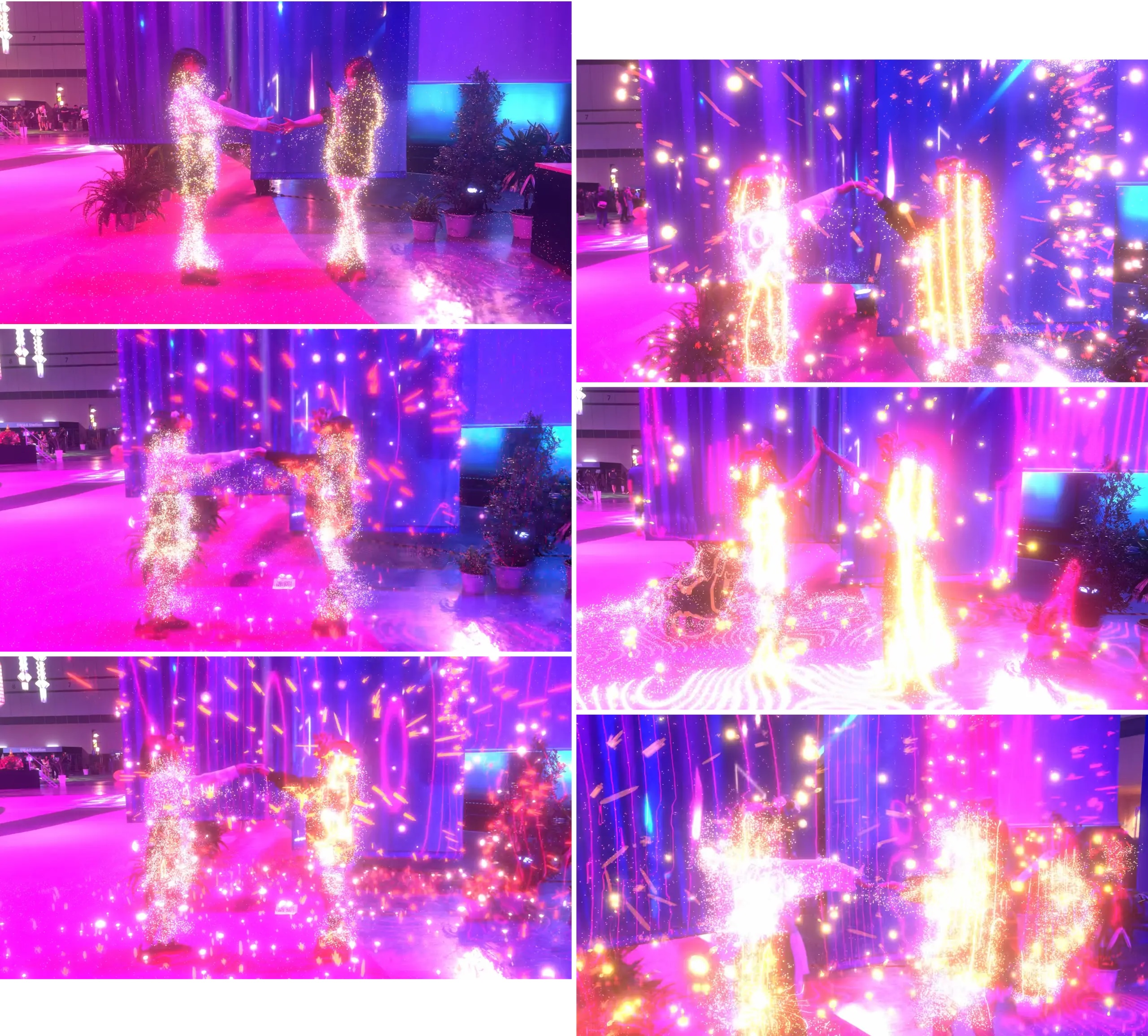}
    \caption{The process of merging two cyberdelic mixed reality experiences}
    \label{fig:merge}
\end{figure}

\subsection{Designing the Interaction: Mixing Realities Through Touch}

The central feature of FungiSync is the entanglement that occurs when participants touch, as Figure \ref{fig:bodycontact} illustrates. When the system detects that two participants' hands are in close proximity—within approximately 10 centimeters, determined by comparing tracked hand positions—a connection state is initiated. During this state, visual elements begin to transfer between the two participants' MR overlays, simulating the resource exchange that occurs when fungal hyphae fuse between plant roots.

The transfer is implemented as a gradual blending process. Each participant's shader receives data about the other's visual layer, and elements from the other's Umwelt begin to render in their own view---initially faint and localized to the region near the touch point, then increasingly vivid and widespread as touch continues, as Figure \ref{fig:merge} shows. The blending is asymmetric: the precise mixture depends on each participant's position and orientation, creating slightly different experiences of the ``shared'' content. This asymmetry models the biological reality that mycorrhizal exchange is not perfectly reciprocal.

When participants release their touch, the transferred elements fade gradually over approximately thirty seconds, leaving a lingering trace of the encounter. This temporal decay models the idea that resources transferred through mycorrhizal networks are consumed over time---without continued exchange, each organism returns to relying on its own resources. Participants who engage in frequent, sustained touches with multiple partners find their visual worlds richly layered with traces of many encounters; those who remain isolated experience a sparser umwelt. This dynamic creates an embodied incentive for connection and a felt consequence for isolation.

FungiSync builds on \citet{hu2023InstantCopresence}'s framework for collocated multiplayer MR, which aligns the spatial poses of all local devices in real time, allowing multiple users to share a unified digital space. Leveraging Apple ARKit's Semantic Person Segmentation, FungiSync identifies nearby human forms and tracks hand positions to detect proximity between participants. Real-time body segmentation ensures that digital artifacts remain appropriately occluded by or positioned around people, preserving spatial realism.

\subsection{Designing the Ritual: Open-Ended Participatory Performance}

\begin{figure}[ht]
    \centering
    \includegraphics[width=0.6\linewidth]{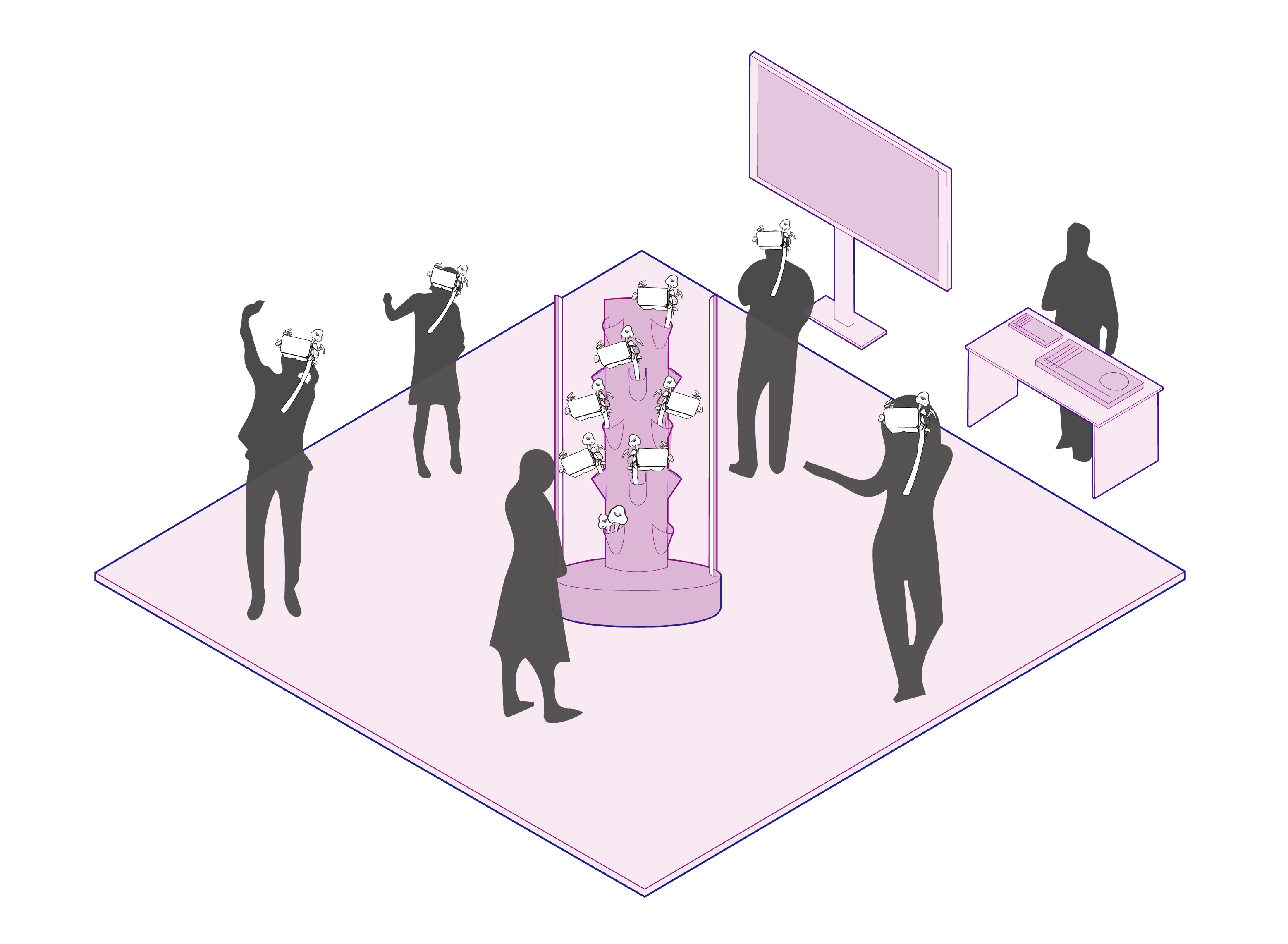}\\
       \includegraphics[width=0.9\linewidth]{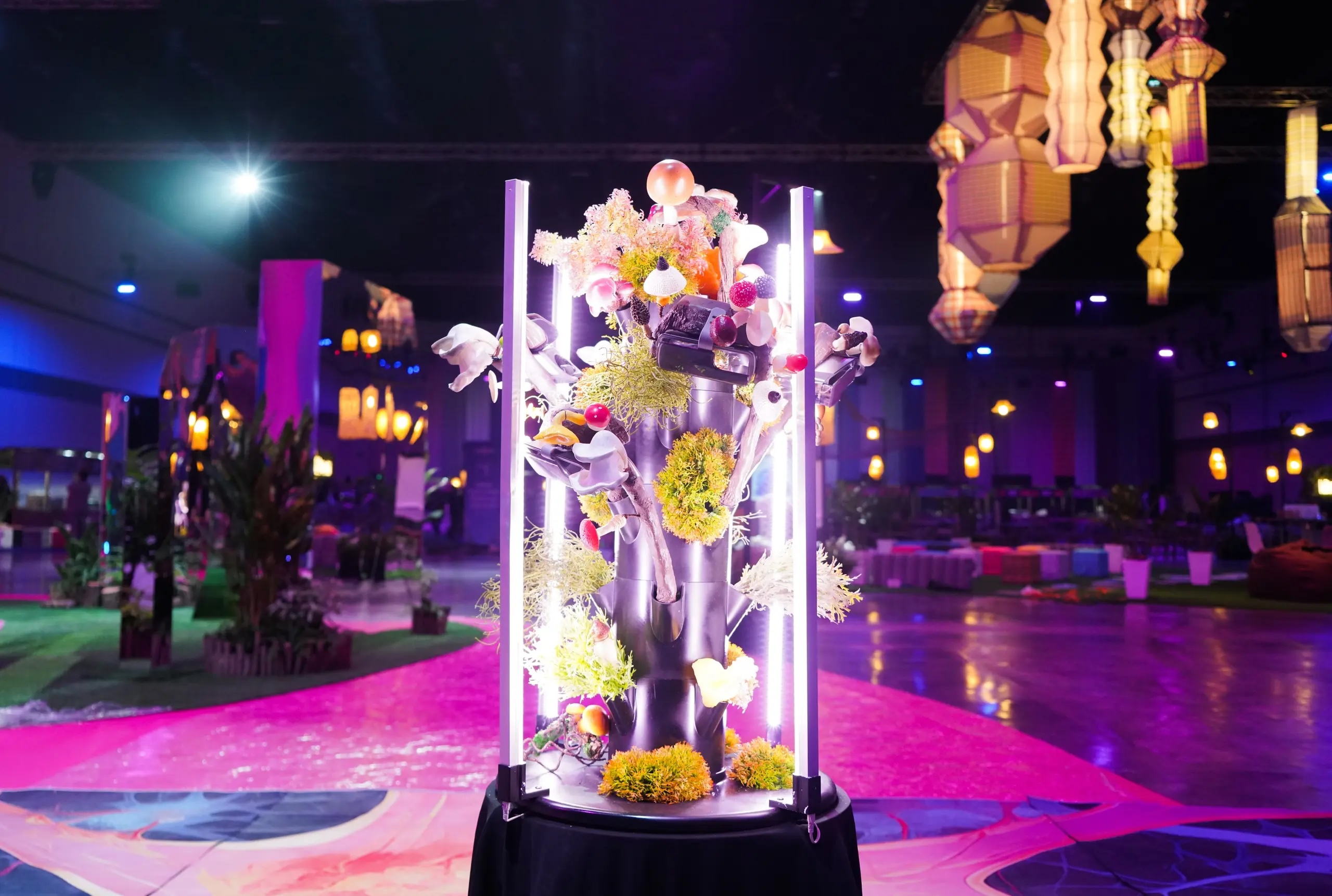}
    \caption{Setup for the Participatory Performance:
The experience is entirely walk-in, walk-out—no rounds required. Visitors can approach the central Substrate Tower at any time to pick up a mask, engage in the performance, and return the mask when finished, creating a fluid, open-ended flow of participation.}
    \label{fig:the-performance-setup}
\end{figure}

FungiSync was designed for exhibition in gallery or festival settings as an open-ended participatory performance, as Figure \ref{fig:the-performance-setup} shows. There is no fixed start or end time; participants can fluidly move between observing and joining. At the center of the space, we place the ``Substrate Tower,'' a sculptural stand resembling a tree stump with slots to hold the MR masks when not in use. This tower serves as both charging dock and ritual focal point: picking up a mask symbolically signifies entering the FungiSync network, and returning it signifies exiting.

Up to six participants can be in the space concurrently, each with their own mask. Spectators without masks can walk around and observe participants, though they cannot see the MR visuals except on a nearby display mirroring each view. We provide a large screen showing a split view of all active perspectives, adding a social dynamic---participants know they are being watched and often perform more intentionally, elevating the ritual quality.

\begin{figure}[ht]
    \centering
    \includegraphics[width=0.45\linewidth]{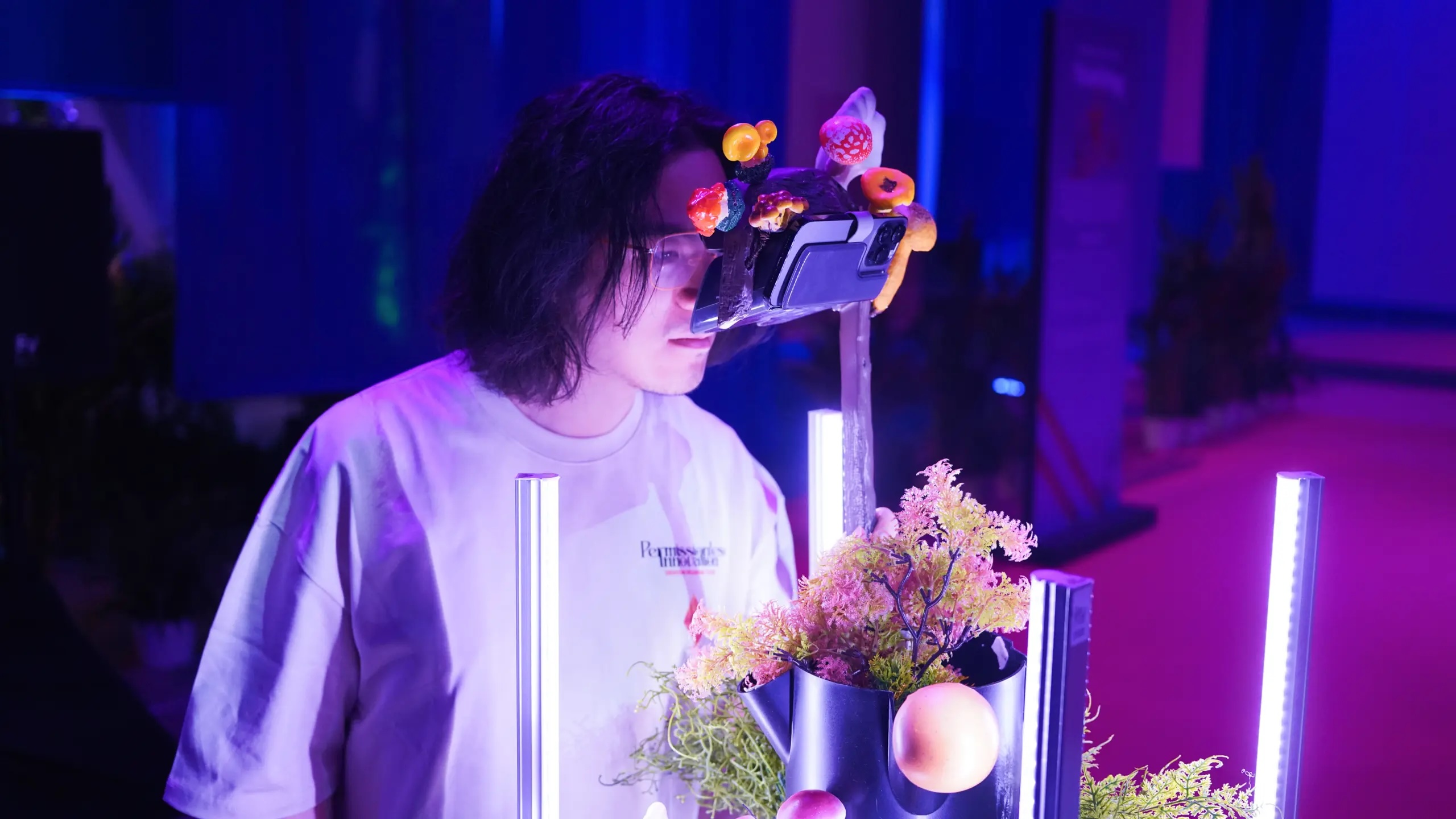}
    \includegraphics[width=0.45\linewidth]{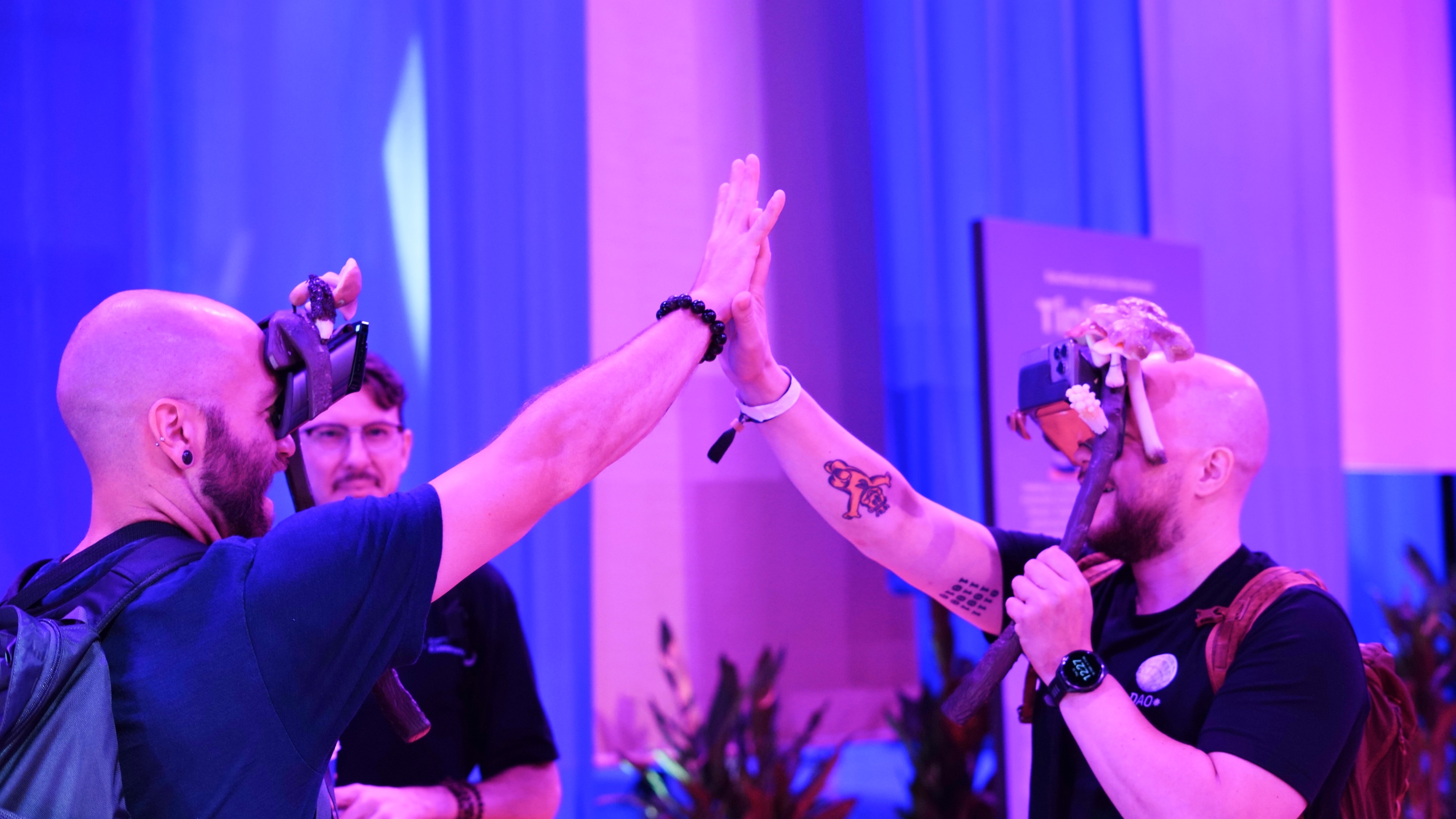}
    \includegraphics[width=0.45\linewidth]{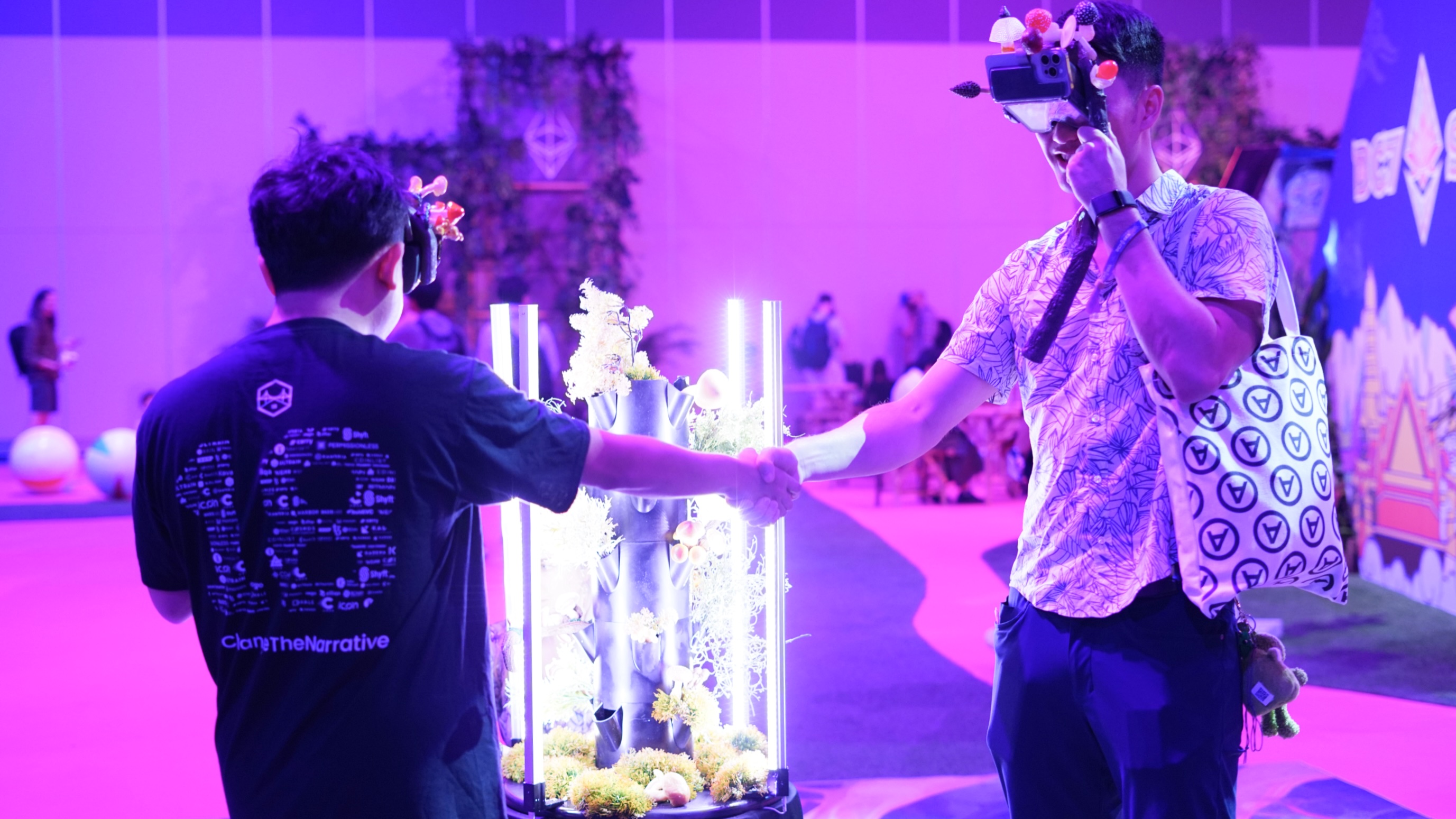}
    \includegraphics[width=0.45\linewidth]{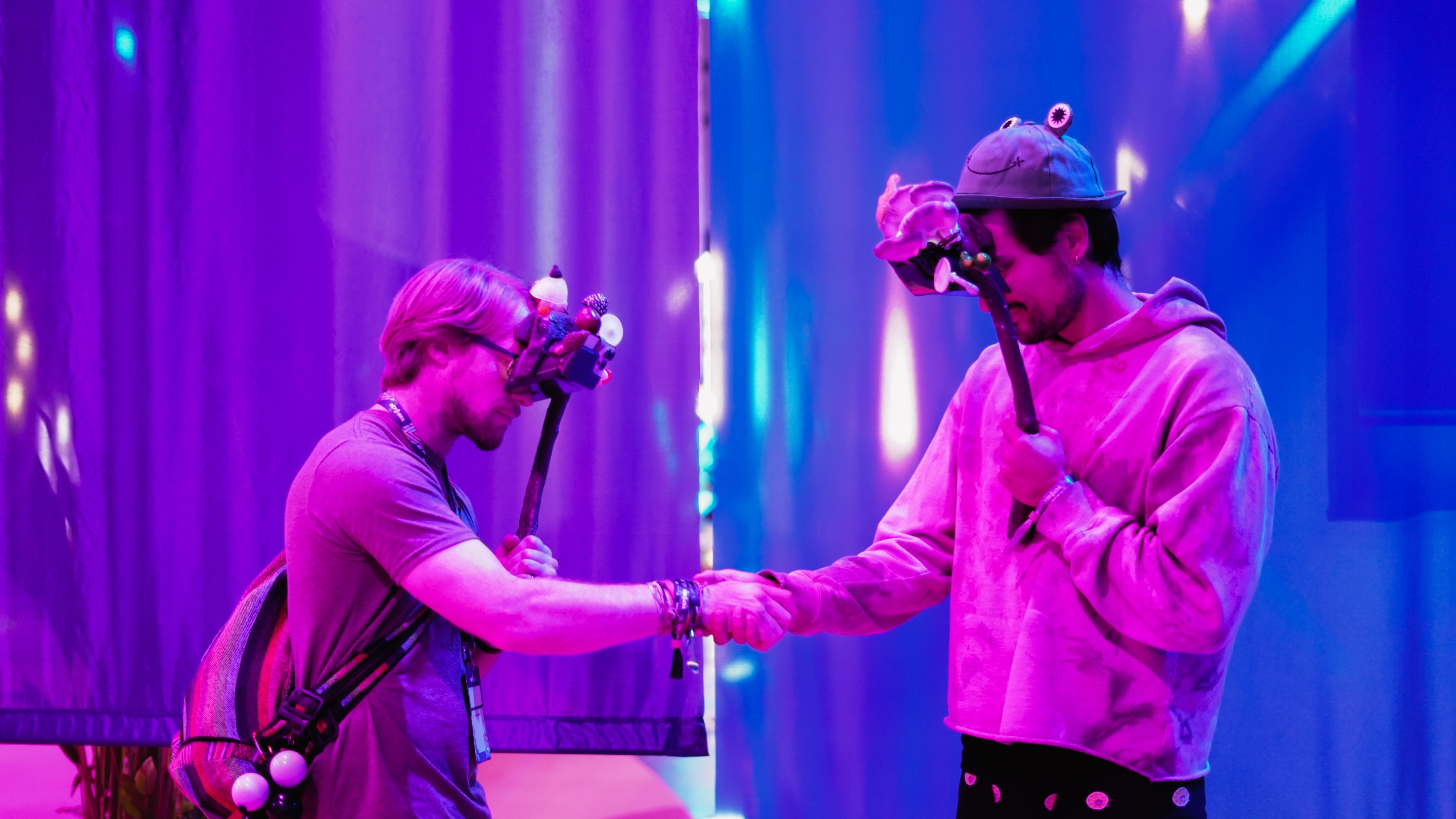}
    \includegraphics[width=0.45\linewidth]{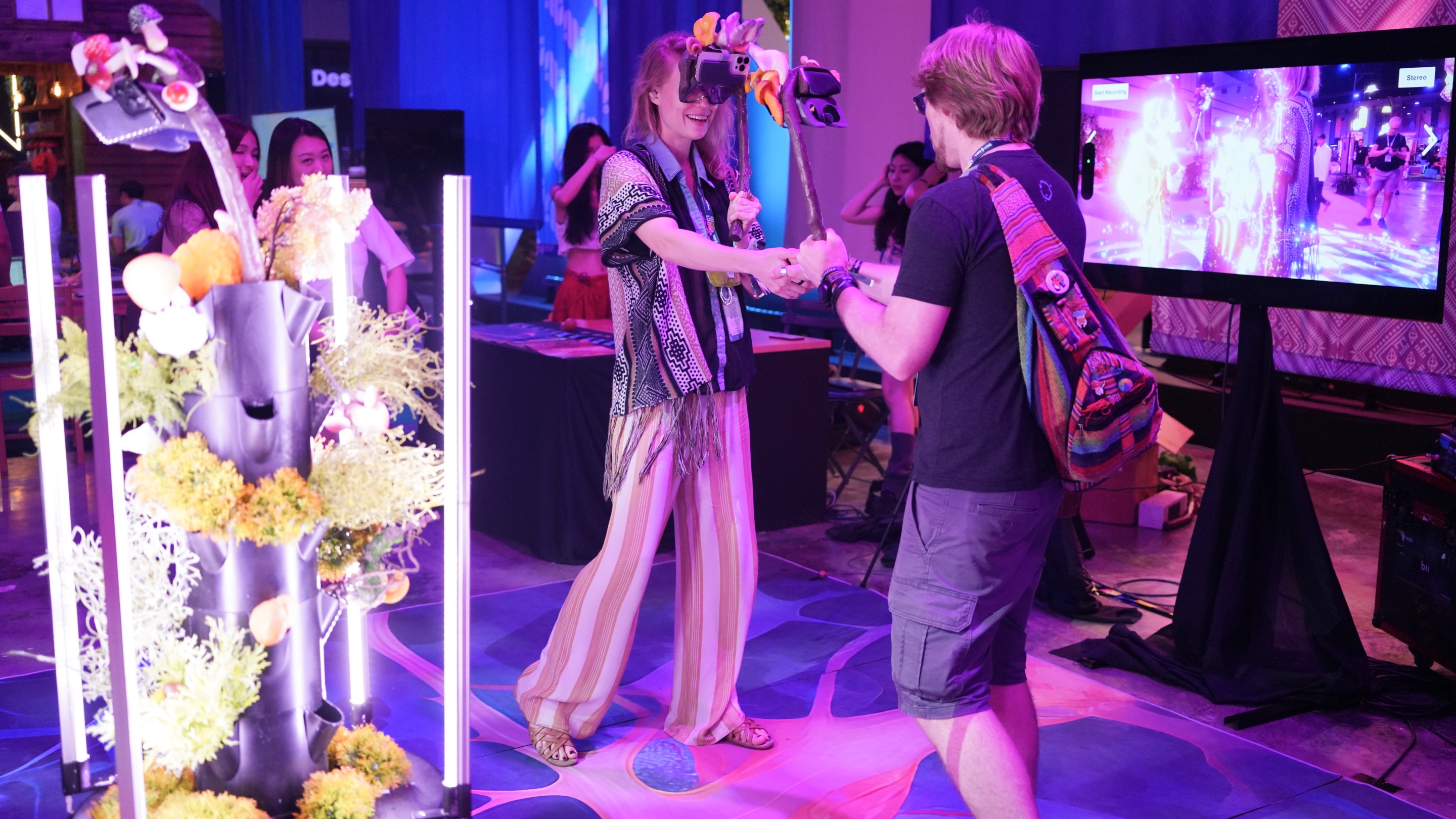}
    \includegraphics[width=0.45\linewidth]{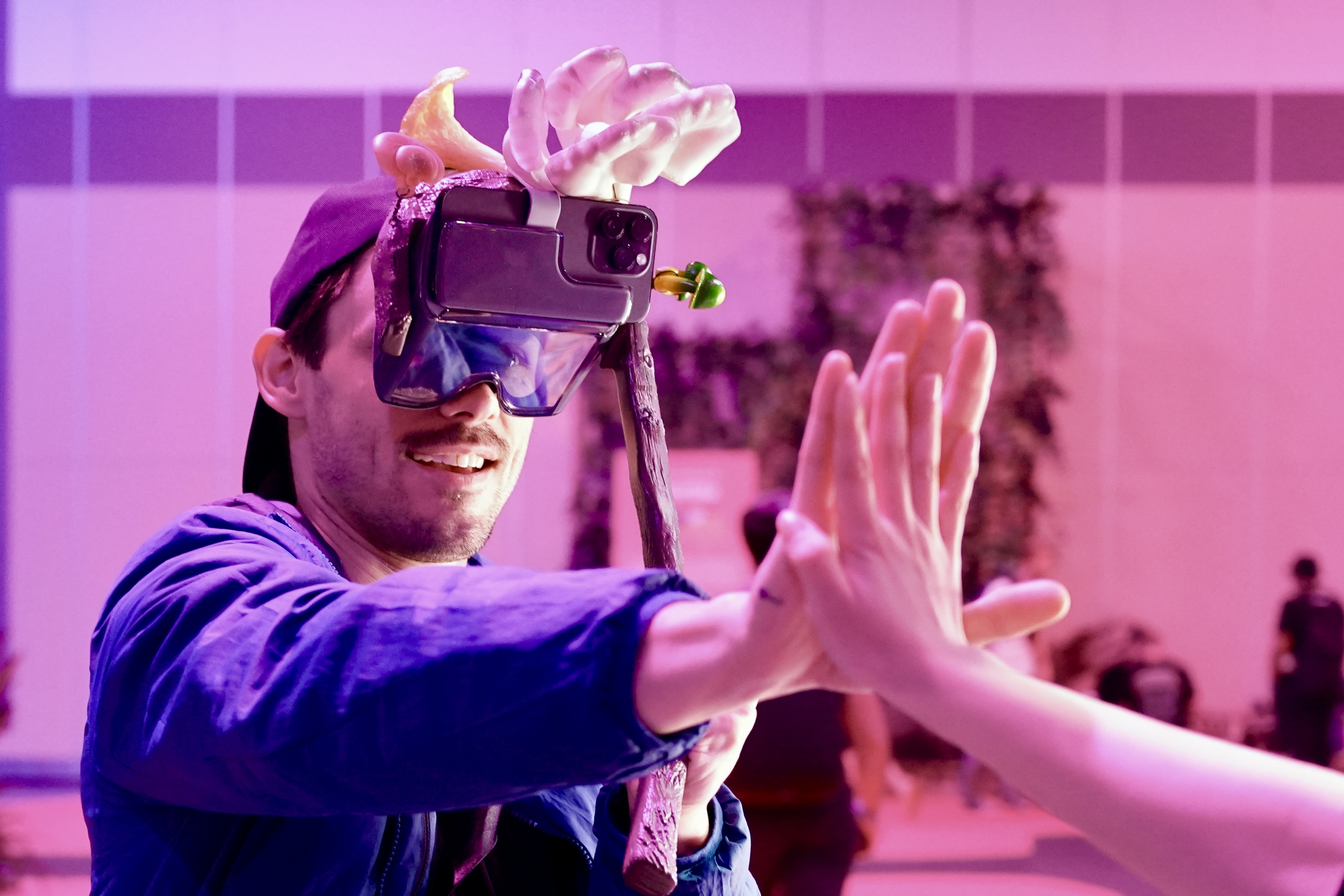}

    \caption{Participants engage in hand contact during the public exhibition, triggering the mixing of their cyberdelic MR overlays.}
    
    \label{fig:twohands}
\end{figure}

\section{Exhibition and Discussion}

FungiSync debuted as a commissioned artwork during a public conference, occupying a dedicated booth in an art gallery setting (see Figure \ref{fig:the-performance-setup}). The installation attracted participants from diverse backgrounds—visual artists, technologists, ecologists, computer scientists, and sociologists—highlighting the interdisciplinary appeal of bio-inspired art. A musician improvised in real time, drawing inspiration from environmental changes and participant interactions to mirror FungiSync's dynamic processes, reinforcing the theme of ecological connectivity through responsive soundscapes.

Post-experience conversations revealed that many participants gained their first visceral understanding of how mycorrhizal networks function to balance resources in forests. Participants expressed strong curiosity about touching each other's hands to experience other realities, often creatively experimenting with different body parts to explore the system's responses (see Figure \ref{fig:twohands}). This playful exploration aligns with soma design principles of non-dualism that blur the boundary between self and other \cite{Hook2018Designingf}.

\subsection{Fungal Epistemics Through Intercorporeal Ritual}

FungiSync posits that fungal epistemics—ways of knowing rooted in mycorrhizal principles of relationality, mutual aid, and networked interdependence—are best accessed through embodied ritual rather than intellectual abstraction alone. Taking seriously the proposition that bodily experience is a legitimate and powerful locus of knowledge production \cite{Shusterman2008Body}, the work frames this more-than-human way of knowing as both an aesthetic experience and an ethical orientation. Just as mycelial networks sustain diverse life forms through reciprocal, polycentric exchange, FungiSync encourages participants to engage in "perspective swapping"—the exchange of augmented realities and embodied actions that mirror fungal resource sharing. Participants do not merely learn about mycorrhizal networks; they enact a stylized version of this exchange. This enactment constitutes a distinct mode of understanding that cultivates a symbiotic relationality. After experiencing FungiSync, the author of \textit{Exploring MycoFi} \cite{jeff2024} reflected on the project's underlying mycelial wisdom, noting how it embodies non-dualistic systems where the blurring of boundaries between "you" and "I" allows participants to deeply affect one another. He described this resonance with decentralized, fluid networks as a form of "underground socialism."

\subsection{Critiquing Digital Individualism}

Beyond establishing more-than-human epistemics, FungiSync offers an implicit critique of the current trajectory of the technology-mediated posthuman, which has largely prioritized personalization and individual optimization. The dominant vision of XR suggests a future where users inhabit tailored perceptual worlds designed to maximize individual preferences and productivity. This extension of algorithmic personalization produces a "posthuman individualism" that resonates with broader concerns regarding digital technology and social fragmentation. As \citet{Turkle2011Alone} has documented, digital devices often engender feelings of isolation even when used for ostensible communication. FungiSync responds to these concerns not by proposing a return to an imagined pre-digital unity, but by visualizing a different technological future: one where digital systems reward connection across difference rather than a retreat into sameness. Rather than isolating users in fully personalized umwelten, we present an experience that connects these distinct perceptual worlds. This approach is informed by the fungal ethos of relational being—where entities are interdependent, and diversity is not only preserved but required for the network to thrive.

\section{Conclusion}

Ultimately, FungiSync translates the mycorrhizal wisdom of interdependence into a felt, embodied experience accessible through mixed reality ritual. By assigning each participant a distinct perceptual world and allowing these worlds to merge through physical touch, the work renders tangible an ecological principle that is otherwise invisible and abstract. The design draws on somaesthetics to treat bodily experience as a site of knowledge and ethics, on ritual theory to frame the handshake as a transformational gesture, and on fungal epistemics to propose relationality as a vital alternative to digital individualism.

\bibliographystyle{ACM-Reference-Format}
\bibliography{reference}

@article{spacal2015connections,
  title={Connections Continium: A Life},
  author={Spacal, Sa{\v{s}}a},
  journal={Experiencing the Unconventional: Science in Art},
  pages={175--206},
  year={2015}
}

@book{Collins2014Interaction,
  title = {Interaction {{Ritual Chains}}},
  author = {Collins, Randall},
  year = 2014,
  series = {Princeton {{Studies}} in {{Cultural Sociology}}},
  publisher = {Princeton University Press},
  address = {Princeton},
  isbn = {978-0-691-12389-9 978-1-4008-5174-4},
  langid = {english}
}

@article{Bayley2025Mycelial,
  title = {Mycelial Matters: Fungal Epistemics and the Birth of the New Materialisms},
  shorttitle = {Mycelial Matters},
  author = {Bayley, Annouchka},
  year = 2025,
  month = oct,
  journal = {Notes and Records: the Royal Society Journal of the History of Science},
  pages = {20250002},
  issn = {1743-0178},
  doi = {10.1098/rsnr.2025.0002},
  urldate = {2026-01-12},
  langid = {english}
}

@article{Cecire2024Mycoaesthetics,
  title = {Mycoaesthetics},
  author = {Cecire, Natalia and Solomon, Samuel},
  year = 2024,
  month = jun,
  journal = {Critical Inquiry},
  volume = {50},
  number = {4},
  pages = {703--724},
  issn = {0093-1896, 1539-7858},
  doi = {10.1086/730345},
  urldate = {2026-01-12},
  langid = {english}
}

@article{Haans2006Mediated,
  title = {Mediated Social Touch: A Review of Current Research and Future Directions},
  shorttitle = {Mediated Social Touch},
  author = {Haans, Antal and IJsselsteijn, Wijnand},
  year = 2006,
  month = jan,
  journal = {Virtual Reality},
  volume = {9},
  number = {2-3},
  pages = {149--159},
  issn = {1359-4338, 1434-9957},
  doi = {10.1007/s10055-005-0014-2},
  urldate = {2026-01-12},
  copyright = {http://www.springer.com/tdm},
  langid = {english}
}

@book{Haraway2016Stayinga,
  title = {Staying with the {{Trouble}}: {{Making Kin}} in the {{Chthulucene}}},
  shorttitle = {Staying with the {{Trouble}}},
  author = {Haraway, Donna J.},
  year = 2016,
  month = aug,
  pages = {dup;9780822373780/1},
  publisher = {Duke University Press},
  doi = {10.1215/9780822373780},
  urldate = {2026-01-12},
  isbn = {978-0-8223-7378-0 978-0-8223-6214-2 978-0-8223-6224-1},
  langid = {english}
}

@book{Hook2018Designingf,
  title = {Designing with the {{Body}}: {{Somaesthetic Interaction Design}}},
  shorttitle = {Designing with the {{Body}}},
  author = {H{\"o}{\"o}k, Kristina},
  year = 2018,
  month = nov,
  publisher = {The MIT Press},
  doi = {10.7551/mitpress/11481.001.0001},
  urldate = {2026-01-12},
  isbn = {978-0-262-34832-4},
  langid = {english}
}

@book{Shildrick2002Embodyingc,
  title = {Embodying the Monster: Encounters with the Vulnerable Self},
  shorttitle = {Embodying the Monster},
  author = {Shildrick, Margrit},
  year = 2002,
  series = {Theory, Culture \& Society},
  edition = {Online-Ausg},
  publisher = {SAGE},
  address = {London},
  isbn = {978-0-7619-7014-9 978-1-4462-2057-3},
  langid = {english}
}

@inproceedings{Ji2025We,
  title = {We {{Are Entanglement}}},
  booktitle = {Proceedings of the {{Special Interest Group}} on {{Computer Graphics}} and {{Interactive Techniques Conference Art Gallery}}},
  author = {Ji, Haru Hyunkyung and Wakefield, Graham},
  year = 2025,
  month = aug,
  pages = {1--3},
  publisher = {ACM},
  address = {Vancouver British Columbia Canada},
  doi = {10.1145/3721249.3731635},
  urldate = {2026-01-12},
  isbn = {979-8-4007-1548-8},
  langid = {english}
}

@article{Yue2025Ritual,
  title = {Ritual Design in Digital Age: {{A}} Comprehensive Analysis of Development and Trend},
  shorttitle = {Ritual Design in Digital Age},
  author = {Yue, Wei and Shen, Jiahui and Hsu, Chun-Cheng},
  year = 2025,
  month = sep,
  journal = {Design Studies},
  volume = {100},
  pages = {101333},
  issn = {0142694X},
  doi = {10.1016/j.destud.2025.101333},
  urldate = {2026-01-12},
  langid = {english}
}

@article{Meltzer1968INTERACTION,
  title = {{{INTERACTION RITUAL}}: {{ESSAYS IN FACE-TO-FACE BEHAVIOR}}. {{By Erving Goffman}}. {{Chicago}}: {{Aldine Publishing Company}}, 1967. 270 Pp. \$5.75},
  shorttitle = {{{INTERACTION RITUAL}}},
  author = {Meltzer, B. N.},
  year = 1968,
  month = sep,
  journal = {Social Forces},
  volume = {47},
  number = {1},
  pages = {110--111},
  issn = {0037-7732, 1534-7605},
  doi = {10.2307/2574751},
  urldate = {2026-01-12},
  langid = {english}
}

@article{Nagel1974Whata,
  title = {What {{Is It Like}} to {{Be}} a {{Bat}}?},
  author = {Nagel, Thomas},
  year = 1974,
  month = oct,
  journal = {The Philosophical Review},
  volume = {83},
  number = {4},
  eprint = {2183914},
  eprinttype = {jstor},
  pages = {435},
  issn = {00318108},
  doi = {10.2307/2183914},
  urldate = {2026-01-12}
}

@book{Pariser2011filter,
  title = {The Filter Bubble: What the {{Internet}} Is Hiding from You},
  shorttitle = {The Filter Bubble},
  author = {Pariser, Eli},
  year = 2011,
  publisher = {Penguin Press},
  address = {New York},
  isbn = {978-1-59420-300-8},
  lccn = {ZA4237 .P37 2011}
}

@book{Shusterman2008Body,
  title = {Body {{Consciousness}}: {{A Philosophy}} of {{Mindfulness}} and {{Somaesthetics}}},
  shorttitle = {Body {{Consciousness}}},
  author = {Shusterman, Richard},
  year = 2008,
  month = jan,
  edition = {1},
  publisher = {Cambridge University Press},
  doi = {10.1017/CBO9780511802829},
  urldate = {2026-01-12},
  copyright = {https://www.cambridge.org/core/terms},
  isbn = {978-0-521-85890-8 978-0-521-67587-1 978-0-511-80282-9}
}

@article{Simard2012Mycorrhizal,
  title = {Mycorrhizal Networks: {{Mechanisms}}, Ecology and Modelling},
  shorttitle = {Mycorrhizal Networks},
  author = {Simard, Suzanne W. and Beiler, Kevin J. and Bingham, Marcus A. and Deslippe, Julie R. and Philip, Leanne J. and Teste, Fran{\c c}ois P.},
  year = 2012,
  month = apr,
  journal = {Fungal Biology Reviews},
  volume = {26},
  number = {1},
  pages = {39--60},
  issn = {17494613},
  doi = {10.1016/j.fbr.2012.01.001},
  urldate = {2026-01-12},
  copyright = {https://www.elsevier.com/tdm/userlicense/1.0/},
  langid = {english}
}

@article{Song2010Interplant,
  title = {Interplant {{Communication}} of {{Tomato Plants}} through {{Underground Common Mycorrhizal Networks}}},
  author = {Song, Yuan Yuan and Zeng, Ren Sen and Xu, Jian Feng and Li, Jun and Shen, Xiang and Yihdego, Woldemariam Gebrehiwot},
  editor = {Van Der Heijden, Marcel},
  year = 2010,
  month = oct,
  journal = {PLoS ONE},
  volume = {5},
  number = {10},
  pages = {e13324},
  issn = {1932-6203},
  doi = {10.1371/journal.pone.0013324},
  urldate = {2026-01-12},
  langid = {english}
}

@book{Tsing2015Mushrooma,
  title = {The {{Mushroom}} at the {{End}} of the {{World}}: {{On}} the {{Possibility}} of {{Life}} in {{Capitalist Ruins}}},
  shorttitle = {The {{Mushroom}} at the {{End}} of the {{World}}},
  author = {Tsing, Anna Lowenhaupt},
  year = 2015,
  month = dec,
  publisher = {Princeton University Press},
  address = {Princeton},
  doi = {10.1515/9781400873548},
  urldate = {2026-01-12},
  isbn = {978-1-4008-7354-8}
}

@book{Turkle2011Alone,
  title = {Alone Together: Why We Expect More from Technology and Less from Each Other},
  shorttitle = {Alone Together},
  author = {Turkle, Sherry},
  year = 2011,
  publisher = {Basic Books},
  address = {New York},
  isbn = {978-0-465-01021-9 978-0-465-02234-2},
  lccn = {HM851 .T86 2011}
}

@book{Uexkull2010Forayb,
  title = {Foray into the Worlds of Animals and Humans: With a Theory of Meaning},
  shorttitle = {Foray into the Worlds of Animals and Humans},
  author = {von Uexk{\"u}ll, Jakob and Sagan, Dorion and {Winthrop-Young}, Goeffrey},
  translator = {O'Neil, Joseph D.},
  year = 2010,
  series = {Posthumanities},
  publisher = {University of Minnesota Press},
  address = {Minneapolis London},
  isbn = {978-0-8166-5900-5 978-0-8166-7538-8 978-0-8166-5899-2}
}

@article{Price2022,
author = {Price, Sara and Bianchi-Berthouze, Nadia and Jewitt, Carey and Steimle, J\"{u}rgen},
title = {Introduction to the Special Issue on Digital Touch: Reshaping Interpersonal Communicative Capacity and Touch Practices},
year = {2022},
issue_date = {June 2022},
publisher = {Association for Computing Machinery},
address = {New York, NY, USA},
volume = {29},
number = {3},
issn = {1073-0516},
url = {https://doi.org/10.1145/3505591},
doi = {10.1145/3505591},
journal = {ACM Trans. Comput.-Hum. Interact.},
month = feb,
articleno = {18},
numpages = {8}
}

@book{jeff2024,
author = {Jeff Emmett},
title = {Exploring MyCoFi - Mycelial Design Patterns for Web3 and Beyond},
year = {2024}
}

@book{merleau2013phenomenology,
  title={Phenomenology of perception},
  author={Merleau-Ponty, Maurice and Landes, Donald and Carman, Taylor and Lefort, Claude},
  year={2013},
  publisher={Routledge}
}

@book{sheldrake2021entangled,
  title={Entangled life: How fungi make our worlds, change our minds \& shape our futures},
  author={Sheldrake, Merlin},
  year={2021},
  publisher={Random House Trade Paperbacks}
}

@inproceedings{smith2022cyberdelics,
  title={Cyberdelics: context engineering psychedelics for altered traits},
  author={Smith, Carl Hayden and Warner, Melissa},
  booktitle={Proceedings of EVA London 2022},
  pages={252--259},
  year={2022},
  organization={BCS Learning \& Development}
}

@inproceedings{hu2023InstantCopresence,
  title = {{{InstantCopresence}}: {{A Spatial Anchor Sharing Methodology}} for {{Co-located Multiplayer Handheld}} and {{Headworn AR}}},
  shorttitle = {{{InstantCopresence}}},
  booktitle = {2023 {{IEEE International Symposium}} on {{Mixed}} and {{Augmented Reality Adjunct}} ({{ISMAR-Adjunct}})},
  author = {Hu, Botao and Zhang, Yuchen and Hao, Sizheng and Tao, Yilan},
  year = {2023},
  month = oct,
  pages = {762--763},
  publisher = {IEEE},
  address = {Sydney, Australia},
  doi = {10.1109/ISMAR-Adjunct60411.2023.00165},
  urldate = {2024-01-03},
  isbn = {9798350328912}
}

@inproceedings{Oliveira2016,
author = {Oliveira, Elen and Bertrand, Philippe and Roel Lesur, Marte and Palomo, Priscila and Demarzo, Marcelo and Cebolla, Ausiàs and Baños, Rosa and Tori, Romero},
year = {2016},
month = {06},
pages = {81-89},
title = {Virtual Body Swap: A New Feasible Tool to Be Explored in Health and Education},
doi = {10.1109/SVR.2016.23}
}

@inproceedings{Hu2024HoloKit,
  author={Botao Amber Hu and Yuchen Zhang and Yilan Elan Tao and Tongzhou Yu},
  booktitle={Companion of the 2024
ACM International Joint Conference on Pervasive and Ubiquitous Computing
Pervasive and Ubiquitous Computing (UbiComp Companion ’24)}, 
  title={HoloKit: Demonstrating an Open-Source Smartphone-Based Mixed Reality
Headset for Mixed Reality Design Education}, 
  year={2024},
  volume={},
  number={},
  doi={10.1145/3675094.3677549}
}

@inproceedings{Stepanova2024Intercorporeal,
author = {Stepanova, Ekaterina R. and Reicke, Bernhard E.},
title = {Intercorporeal Design: Dissolving Self-Other Dualism in Interaction Design},
year = {2024},
isbn = {9798400710421},
publisher = {Association for Computing Machinery},
address = {New York, NY, USA},
url = {https://doi.org/10.1145/3686169.3686206},
doi = {10.1145/3686169.3686206},
booktitle = {Proceedings of the Halfway to the Future Symposium},
articleno = {5},
numpages = {9},
location = {Santa Cruz, CA, USA},
series = {HttF '24}
}

@article{lin2024cell,
  title={Cell space: Augmented awareness of intercorporeality},
  author={Lin, Rem RunGu and Hu, Botao Amber and Ke, Koo Yongen and Wu, Wei and Zhang, Kang},
  journal={Proceedings of the ACM on Computer Graphics and Interactive Techniques},
  volume={7},
  number={4},
  pages={1--10},
  year={2024},
  publisher={ACM New York, NY, USA}
}

@incollection{desnoyers2020body,
  title={Body RemiXer: extending bodies to stimulate social connection in an immersive installation},
  author={Desnoyers-Stewart, John and Stepanova, Ekaterina R and Riecke, Bernhard E and Pennefather, Patrick},
  booktitle={ACM SIGGRAPH 2020 Art Gallery},
  pages={394--400},
  year={2020}
}

\end{document}